\documentclass[twocolumn,preprintnumbers,amsmath,amssymb,showpacs]{revtex4}
\usepackage{dcolumn}
\usepackage{color}
\usepackage{bm}
\usepackage{graphicx}
\usepackage{epsfig}
\usepackage{amsmath}
\usepackage{amsfonts}
\usepackage{amssymb}

\def \e{\mathrm{e}}
\def \half{{1 \over 2}}
\def \pxipy{$p_x\!+\!ip_y$ } 
\def \He3A{$^3$HeA} 
\def \SRO{{Sr$_2$RuO$_4$}} 
\def \xvec{\boldsymbol{x}} 
\def \omgvec{\boldsymbol{\Omega}} 
\def \Do2eF{\Delta_{0\phantom{2}}^{\phantom{0}2}/\epsilon_F} 
\def \eC{\epsilon_c} 
\def \eF{\epsilon_F} 
\def \Rxi{(R/\xi_p)^2} 
\def \xiR{(\xi_p/R)^2} 
\def \w{W_0} 
\def \ww{W_0^{\phantom{0}2}} 
\def \lL{l_\Lambda} 
\def \tE{\theta_E} 
\def \Swave{{\it s}-wave } 
\def \etal{{\it et al.}~} 

\begin{document}
\draft

\newcommand{\Wj}[6] 
    { \left( \begin{array}{ccc} #1 & #2 & #3 \\ #4 & #5 & #6 \end{array} \right) }

\title{
    Majorana fermions of a two-dimensional \pxipy superconductor}
\author{
    Yaacov E.~Kraus$^{1}$, Assa Auerbach$^{2}$,
    H.A.~Fertig$^{3}$ and Steven H.~Simon$^{4}$ }
\affiliation{
    1) Department of Condensed Matter Physics, Weizmann Institute of Science, Rehovot 76100,
       Israel\\
    2) Department of Physics, Technion, Haifa 32000, Israel\\
    3) Department of Physics, Indiana University, Bloomington, IN 47405, USA\\
    4) Rudolf Peierls Centre for Theoretical Physics, University of Oxford,
       Oxford OX1 3NP, UK}

\begin{abstract}
To investigate Majorana fermionic excitations of a \pxipy
superconductor, the Bogoliubov-de-Gennes equation is solved on a
sphere for two cases: (i) a vortex-antivortex pair at opposite poles
and (ii) an edge near the south pole and an antivortex at the north
pole. The vortex cores support a state of two Majorana fermions, the
energy of which decreases exponentially with the radius of the
sphere, independently of a moderate disorder potential. The
tunneling conductance of an electron into the superconductor near
the position of a vortex is computed for finite temperature, and is
compared to the case of an \Swave superconductor. The zero bias
conductance peak of the antivortex is half that of the vortex. This
effect can be used as a probe of the order parameter symmetry, and
as a direct measurement of the Majorana fermion.
\end{abstract}

\maketitle


\section{Introduction}
\label{Sec:Intro} %
In the last few years it has appeared increasingly likely that
nontrivial (or non-Abelian) topological phases of matter
\cite{NayakRMP} may be produced in the laboratory. In fact, it is
quite possible that these phases of matter have been produced,
although current experiments still leave room for doubt. Interest in
such phases of matter is driven to a large extent by their possible
application for building naturally error resistant, so-called
``topological" quantum computers \cite{NayakRMP}. Among such
topological phases of matter, perhaps the simplest is of the
``Ising" or $SU(2)_2$ class \cite{NayakRMP}, which correspond to
chiral \pxipy BCS paired superconductors \cite{Bravyi}.

There are several possible physical systems where \pxipy pairing is
believed to be realized, including the A phase of superfluid $^3$He
\cite{He3Volovik} (\He3A), the exotic superconductor \SRO
\cite{RiceSigrist}, and the $\nu = 5/2$ quantum Hall state
\cite{MooreRead,Grieter}. In addition, there have been recent
proposals to realize \pxipy pairing in cold fermion gases
\cite{GurarieRadzihosky}. For quantum information processing
applications, two-dimensionality (or at least
quasi-two-dimensionality) is necessary. This is certainly the case
in quantum Hall systems, and may also be achievable for \SRO \
(which is a layered structure), for \He3A films, and also
potentially in cold atomic systems. For the purpose of this paper we
will assume quasi-two-dimensionality, although some parts of our
results are more general.

In these (weak coupling) \pxipy systems, certain types of vortices
(quasiparticles in the quantum Hall context \cite{ReadGreen}) are
believed to carry zero energy Majorana fermions
\cite{Volovik,ReadGreen}, which are topologically protected degrees
of freedom. In \SRO \ and \He3A the vortices that carry the Majorana
fermions are the so-called half-quantum vortices, which can be
thought of as a vortex in the order parameter of one spin species,
without a vortex of the opposite species \cite{TewariKim}. (Note
that in spin-polarized \pxipy systems, including proposed atomic gas
realizations or the 5/2 state, there is no half-quantum vortex and
the full quantum vortex carries the Majorana fermion).

A Majorana fermion is an operator which satisfies the fermionic
anti-commutation relation $\{ \eta^\dag(\xvec), \eta(\xvec') \} =
2\delta(\xvec - \xvec')$, but equals to its own hermitian conjugate
$\eta \dag = \eta$. Therefore $\eta^2 = \eta ^{\dag 2} = \eta^\dag
\eta = 1$. A fermion occupation number state can be defined as a
linear combination of two Majorana fermions localized in two
distinctive vortices $\psi(\xvec) = ( \eta_i(\xvec) + i\eta_j(\xvec)
)/\sqrt{2}$ and $\psi^\dag(\xvec) = ( \eta_i(\xvec) - i\eta_j(\xvec)
)/\sqrt{2}$. This operator satisfies the usual fermionic relations
$\{ \psi^\dag(\xvec), \psi(\xvec') \} = \delta(\xvec - \xvec')$ and
$\psi^{\dag2} = \psi^2 = 0$. We shall call such a fermionic
occupation number state a Majorana state. The energy of the Majorana
state reflects the exponentially small hybridization between the
wavefunctions of the two localized Majorana fermions.

The zero energy of the Majorana fermion in a single vortex is
believed to be topologically protected against weak disorder
\cite{ReadGreen}. But in the case of more than one vortex, an
experimentally relevant question is whether the exponential
localization and hybridization of the Majorana state is a property
only of a clean system, or is it robust against the inclusion of
disorder. We find that these properties survive even in the presence
of a moderate disorder.

Another unique property of the \pxipy order parameter is the
existence of low energy chiral states, which are localized along the
edge of the sample \cite{ReadGreen}. If a single vortex is present,
the Majorana state is split between the vortex core and the edge.

Let us suppose that in one of the above systems, the relevant
Majorana-fermion-carrying vortex has been created
\cite{e4quantumHall}. An important next step would be to design an
experiment to demonstrate that the Majorana fermion is present in
such a vortex \cite{DemlerTewari}. In the case of \SRO, one obvious
experiment would be an energy-resolved tunneling experiment, which
measures the local density of states (LDOS) \cite{Shore}. An
observation of a localized mode at precisely zero energy would be
direct evidence of the Majorana fermion. For cold atoms, an
analogous experiment for observing the LDOS would be an
energy-resolved local particle annihilation experiment. For the
other realizations of \pxipy order it is not as clear how such an
experiment would be performed \cite{tunneling}.

In principle such tunneling experiments could provide definitive
evidence for the Majorana fermion. However, in practice they may be
prohibitively difficult. In the vortex, there will exist sub-gap
bound states in the core known as Caroli-de-Gennes-Matricon (CdGM)
states \cite{CdGM,KopninSalomaa}. The spacing between the CdGM
states is approximately $\eC = \Do2eF$, where $\Delta_0$ is the gap
(presumably on order of the critical temperature) and $\eF$ is the
Fermi energy. Since the experimentally observed tunneling spectrum
will be smeared by the temperature, this tunneling experiment would
naively only have a clear signature for $T < \eC$. Unfortunately
such low temperatures could potentially be unattainable in any of
the proposed realizations ($\eC \approx 7 \mu K$ in \He3A, and $ <
0.1 mK$ in \SRO).

We find that within the reachable temperature region, $\eC < T <
\Delta_0$, the central peak of the smeared LDOS of the antivortex is
half the height of the peak of the vortex. We shall see that this
distinction is clear evidence of the \pxipy symmetry of the order
parameter, and of the existence of the Majorana fermion.

Generally speaking, a physical asymmetry between a vortex and an
antivortex can occur only in superconductors which break time
reversal symmetry, such as \pxipy superconductors \cite{Luke}. The
order parameter of such superconductors involves internal angular
momentum, which is interlaced with the angular momentum of the
vortex according to theirs relative directions \cite{Tanaka}.

This paper is organized as follows:

In Sec.~\ref{Sec:BdG} we implement the \pxipy superconductor on a
sphere, with vortex-antivortex pair at the poles \cite{Assa}. Using
monopole harmonics functions, we numerically solve the Bogoliubov
de-Gennes (BdG) equation, and get the full BdG spectrum.

In Sec.~\ref{Sec:Disorder} we test the exponential decay of the
Majorana state energy as a function of the distance between the
vortices in the presence of disorder. We find it is unaffected, even
in the presence of a moderate disorder.

In Sec.~\ref{Sec:Edge} we put an edge around the south pole of the
sphere, and observe the edge excitations and their linear
dispersion.

In Sec.~\ref{Sec:LDOS} we calculate the tunneling conductance of an
electron into the superconductor near the position of a vortex or an
antivortex at zero temperature and at elevated temperature. We find
an asymmetry effect in the zero bias conductance between the vortex
and the antivortex, which one can use as ``smoking gun'' evidence of
the existence of the Majorana fermion. We compare it to the
tunneling spectrum for a regular \Swave superconductor, to support
this conclusion. Our analysis shows that this effect will occur for
any spin polarized chiral superconductor (chiral-{\it p},
chiral-{\it d}, \ldots). However, it occurs for the single vortex
only for the chiral-{\it p} case, whereas it occurs for the double
vortex for chiral-{\it d}, and correspondingly higher vortices for
higher pairing symmetries. Some of the results of this paper were
recently published in a short format \cite{PRL}.


\section{$\textrm{BdG}$ theory on a sphere}
\label{Sec:BdG}
Consider a two-dimensional uniform \pxipy superconductor of spinless
fermions. The excitation spectrum is given by the BdG equation
\cite{BdG}
\begin{equation} \label{Eq:BdG}
    \left( \begin{array}{cc}
           \widehat{T} - \eF + W      & \Delta \\
           \Delta^{\dagger}            & -(\widehat{T} - \eF + W) \end{array} \right)
    \left( \begin{array}{c} u_n \\ v_n \end{array} \right)
    = E_n \left( \begin{array}{c} u_n \\ v_n \end{array} \right),
\end{equation}
where $\widehat{T}$ is the kinetic energy operator, and $\eF$ is the
Fermi energy. $W$ denotes electrostatic potential, and will be zero
in this section. $\Delta$ is the order parameter field, which
according to the \pxipy symmetry is of the form
\begin{eqnarray} \label{Eq:Delta}
    \Delta (\xvec - \xvec')
        & = & \Delta_0 \frac{\partial_x + i\partial_y}{ik_F} \frac{1}{4\pi\xi_p^{\phantom{p}2}}
              \e^{-\frac{(\xvec - \xvec')^2}{4\xi_p^{\phantom{p}2}}} \\
        & = & \frac{\Delta_0}{8\pi i \xi_p^{\phantom{p}4} k_F} [(x-x') + i(y-y')]
              \e^{-\frac{(\xvec - \xvec')^2}{4\xi_p^{\phantom{p}2}}}.    \nonumber
\end{eqnarray}
Here $\Delta_0$ is the pairing amplitude, and $k_F$ is the Fermi
wavevector, given by $\eF = k_F^{\phantom{w}2}/2m^*$, where $m^*$ is
the electron effective mass. $\xi_p$ is the pairing range, which is
usually taken to be zero for simplicity, whereas in quantum Hall
systems it is comparable to the magnetic length. In Fourier space
\begin{equation}  \label{Eq:Dk}
    \Delta_{\bf k} = \Delta_0 \frac{k_x + ik_y}{k_F} \e^{-k^2 \xi_p^{\phantom{p}2}},
\end{equation}
which shows that $\Delta_0$ is approximately the energy gap.
Although we have written down a special case of \pxipy order
parameter, the ${\bf k} \rightarrow 0$ part is universal.

All the calculations in this paper are considered to be in the short
range pairing limit, where $k_F\xi_p \ll 1$. However, the form
Eq.~\ref{Eq:Dk} remains acceptable up to $k_F\xi_p \approx 1$,
although due to the exponential factor the amplitude is highly
reduced (but can be compensated by multiplying by a factor of
$\e^{k_F^{\phantom{w}2} \xi_p^{\phantom{p}2}}$). We have checked
that none of our results change substantially even for $k_F\xi_p
\approx 1$.

A vortex $(+)$ and an antivortex $(-)$ are described by the order
parameters
\begin{equation} \label{Eq:Delta_pm}
    \Delta_{\pm}(\xvec,\xvec') = \Delta(\xvec-\xvec')
                                 f_{\rm v}(\bar{r}/\xi) e^{ \pm i\bar{\phi}},
\end{equation}
where $\bar{r}, \bar{\phi}$ denote the polar coordinates of the pair
center of mass $(\xvec+\xvec')/2$. The amplitude profile of the
vortex $f_{\rm v}(x)$ vanishes at the origin, and approaches unity
at \mbox{$x \gg 1$}. $\xi$ is Pippard's coherence length
\cite{Tinkham}
\begin{equation}  \label{Eq:xi}
    \xi = \frac{2 \eF} {\pi \Delta_0 k_F}.
\end{equation}
Note that for the vortex $\Delta_+$, the relative and the center of
mass angular momenta are aligned, while for the antivortex
$\Delta_-$, they have opposite chirality.

\begin{figure}[htb]
\vspace{-0.3cm}
\begin{center}
\includegraphics[width=5cm,angle=0]{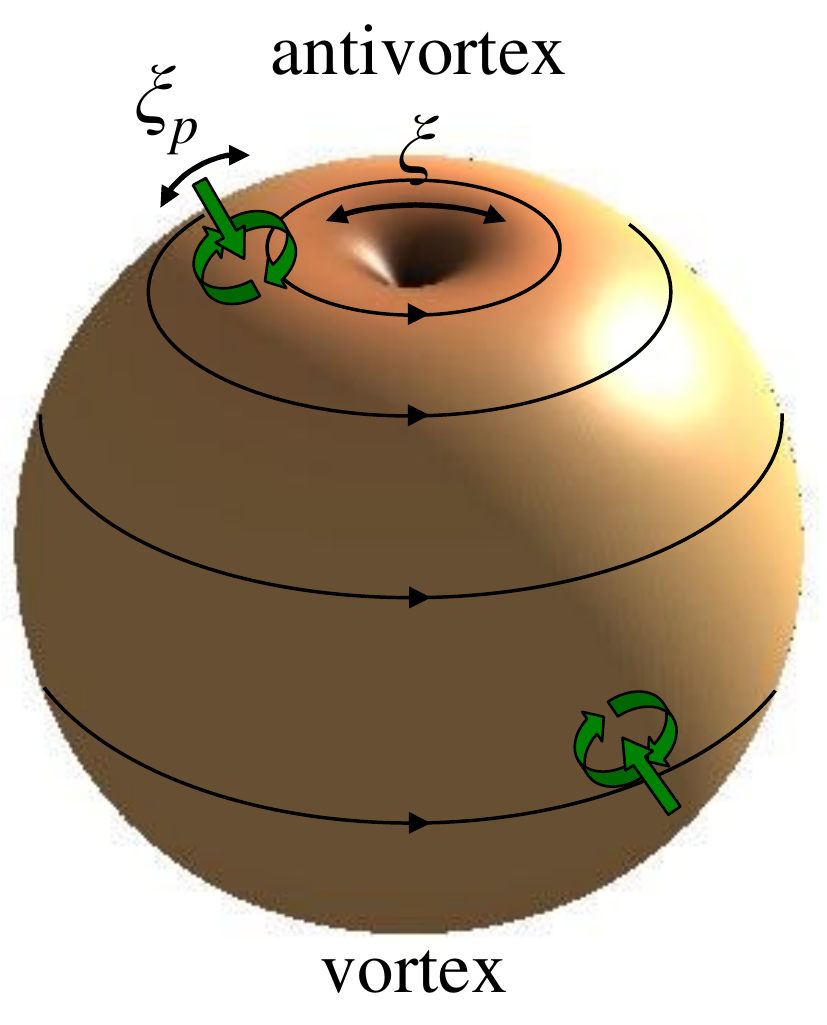}
\caption{  \label{Fig:sphere} %
A vortex-antivortex pair of the \pxipy superconductor on the sphere,
described by Eq.~(\ref{Eq:Delta_V}). Thin black lines represent the
current flow. Wide green arrows represent the pair relative angular
momentum. $\xi_p$ is the pairing range. $\xi$ is the coherence
length, which determines the vortex core size. }
\end{center}
\end{figure}

We implement the BdG equation on a sphere of radius $R$,
parameterized by the unit vector $\omgvec = (\theta,\phi)$. The
spherical geometry has two important advantages: (i) It has no
boundaries, which strongly affect the low energy spectrum (as will
be discussed in Sec.~\ref{Sec:Edge}). (ii) It enables the use of
monopole harmonics functions as a basis, which appears to be very
convenient for the \pxipy pairing. However, the spherical symmetry
forces us to consider an antipodal vortex-antivortex pair. We set
such a vortex-antivortex pair in the north and south poles,
respectively, see Fig.~\ref{Fig:sphere}. The azimuthal symmetry of
this configuration conserves the azimuthal angular momentum, which
greatly reduces the computational difficulty of the BdG
diagonalization.

The spherical geometry is used in this paper for calculations on
finite size geometry, where the physical limit of far separated
vortex-antivortex pair in the two-dimensional plane is approached in
the $R \to \infty$ limit. Since the quantities we are interested in
will be found to decay exponentially fast with $R/\xi$, our finite
sphere calculations will be relevant at values of $R/\xi$ which are
not enormously large, i.e. in moderate vortices density. Of course,
our calculations would be even more directly relevant to thin
spherical shells which might be experimentally created on small
spherical substrates.

The order parameter field on the sphere is taken to be of the
following form \cite{Moller}
\begin{eqnarray}
    \Delta_V (\omgvec,\omgvec')
        & = & \Delta_p (\omgvec,\omgvec') F_V(\bar{\omgvec}),  \label{Eq:Delta_V} \\
    \Delta_p (\omgvec,\omgvec')
        & = & \frac{\Delta_0} {(4\pi\xi_p^{\phantom{p}2}) (l_F + \half)} \nonumber\\
        &   & \times (\alpha \beta' - \beta \alpha')
                     | \alpha \alpha'^* + \beta \beta'^* |^{ 2\Rxi }, \label{Eq:Delta_p}\\
    \alpha & = & \cos(\theta/2), \nonumber\\
    \beta  & = & \sin(\theta/2) \e^{-i\phi},  \label{Eq:ab}
\end{eqnarray}
where $\Delta_p$ gives the pairing of the particles, and is
constructed by the spinor functions $\alpha$ and $\beta$. The
$(\ldots)$ factor in $\Delta_p$ acquires a $2\pi$ phase winding when
$\omgvec$ encircles $\omgvec'$, which describes \pxipy pairing. The
$|\ldots|$ factor keeps the particles within relative distance
$\xi_p$. $l_F$ is the Fermi angular momentum, given by
\begin{equation}  \label{Eq:eF}
    \eF = \frac{l_F (l_F+1)}{2m R^2}.
\end{equation}
In this way $\Delta_p$ reproduces $\Delta$ (Eq.~\ref{Eq:Delta}) in
the large $R$ limit.

$F_V(\bar{\omgvec})$ describes the vorticity of the pair center of
mass $\bar{\omgvec} = (\omgvec + \omgvec')/2$. We choose $F_V$ to
describe an antivortex on the north pole and a vortex on the south
pole, depicted in Fig.~\ref{Fig:sphere}. For the vortex pair field,
we use (without self consistency) the approximate solution of the
Gross-Pita¨evskii equation of a vortex \cite{PS}
\begin{eqnarray}   \label{Eq:F_V}
    F_V(\omgvec)
        & = & \frac{\sin\theta \cdot R/\xi} {\sqrt{ 1 + (\sin\theta \cdot R/\xi)^2 } } \e^{i\phi},
\end{eqnarray}
with $\xi_p < \xi \ll R$ for simplicity.

We expand the order parameter as a series of monopole harmonics
\cite{WuYang} $Y_{q l m}$, of $q=-\half$ and $l,m$ half integers
\cite{ReadGreen,Moller}. These monopole harmonics represent
eigenstates of a particle on a sphere in a radial magnetic field
with \mbox{$2q = -1$} flux quanta penetrating into the sphere (the
$-$ sign was chosen for correspondence with the composite fermion
picture of the $\nu=5/2$ state). In this basis, the BdG equation is
represented as a matrix
\begin{eqnarray} \label{Eq:BdGlmlm}
    \left( \begin{array}{cc}
           T_{lm,l'm'}      & \Delta^V_{lm,\bar{l}'\bar{m}'} \\
           (\Delta^{V\dag})_{\bar{l}\bar{m},l'm'}   & -T_{\bar{l}\bar{m},\bar{l}'\bar{m}'} \end{array} \right)
    \left( \begin{array}{c} u_{n,l'm'} \\ v_{n,\bar{l}'\bar{m}'} \end{array} \right) \nonumber \\
    = E_n \left( \begin{array}{c} u_{n,lm} \\ v_{n,\bar{l}\bar{m}} \end{array} \right),
\end{eqnarray}
with summation over primed indices, and with the following matrix
elements:
\begin{eqnarray}
    T_{lm,l'm'} & = & \delta_{ll'} \delta_{mm'} \; \eF \left(
                      \frac{l(l + 1) - {1 \over 4}} {l_F(l_F + 1) - {1 \over 4}} - 1
                      \right), \label{Eq:Tlmlm} \\
    \Delta^V_{lm,l'm'} & = & \delta_{m', 1-m} \Delta_0 \sqrt{ {\textstyle {1 \over 16\pi} }
                           (2l + 1) (2l' + 1) }  \label{Eq:DVlmlm} \\
                     &   & \times \left( D_l + (-1)^{l + l'} D_{l'} \right)
                            \sum_{L} \sqrt{ 2L + 1 } \; f^V_L  \nonumber\\
                     &    & \qquad \times \Wj{l}{l'}{L}{\half}{-\half}{0}
                                          \Wj{l}{l'}{L}{-m}{m-1}{1},  \nonumber\\
    D_l & \approx & \frac{l}{l_F} \e^{ -l^2\xiR }.  \label{Eq:Dl}
\end{eqnarray}
The matrix element $\Delta^V_{lm,l'm'}$
(Eqs.~\ref{Eq:DVlmlm}-\ref{Eq:Dl}) is justified in Appendix
\ref{App:DVlmlm}. (A definition for $f^V_L$ can also be found
there). As expected by the azimuthal symmetry, $m$ is a good quantum
number. This dramatically simplifies the numerics, since the BdG
matrix can be diagonalized for each $m$ separately.

Diagonalizing Eq.~(\ref{Eq:BdGlmlm}) for each $m$ produces a set of
energies $E_{n,m}$ and corresponding eigenvectors $u_{n,lm},
v_{n,lm}$. The resultant BdG wavefunctions on the sphere are
\begin{eqnarray}
    u_{n,m}(\omgvec) & = & \sum_{l} u_{n,lm} Y_{-\half,l,m}(\omgvec)  \label{Eq:un} \\
    v_{n,m}(\omgvec) & = & \sum_{l} v_{n,lm} Y_{-\half,l,-m+1}^*(\omgvec).  \label{Eq:vn}
\end{eqnarray}

\begin{figure}[htb]
\begin{center}
\includegraphics[width=8cm,height=5cm,angle=0]{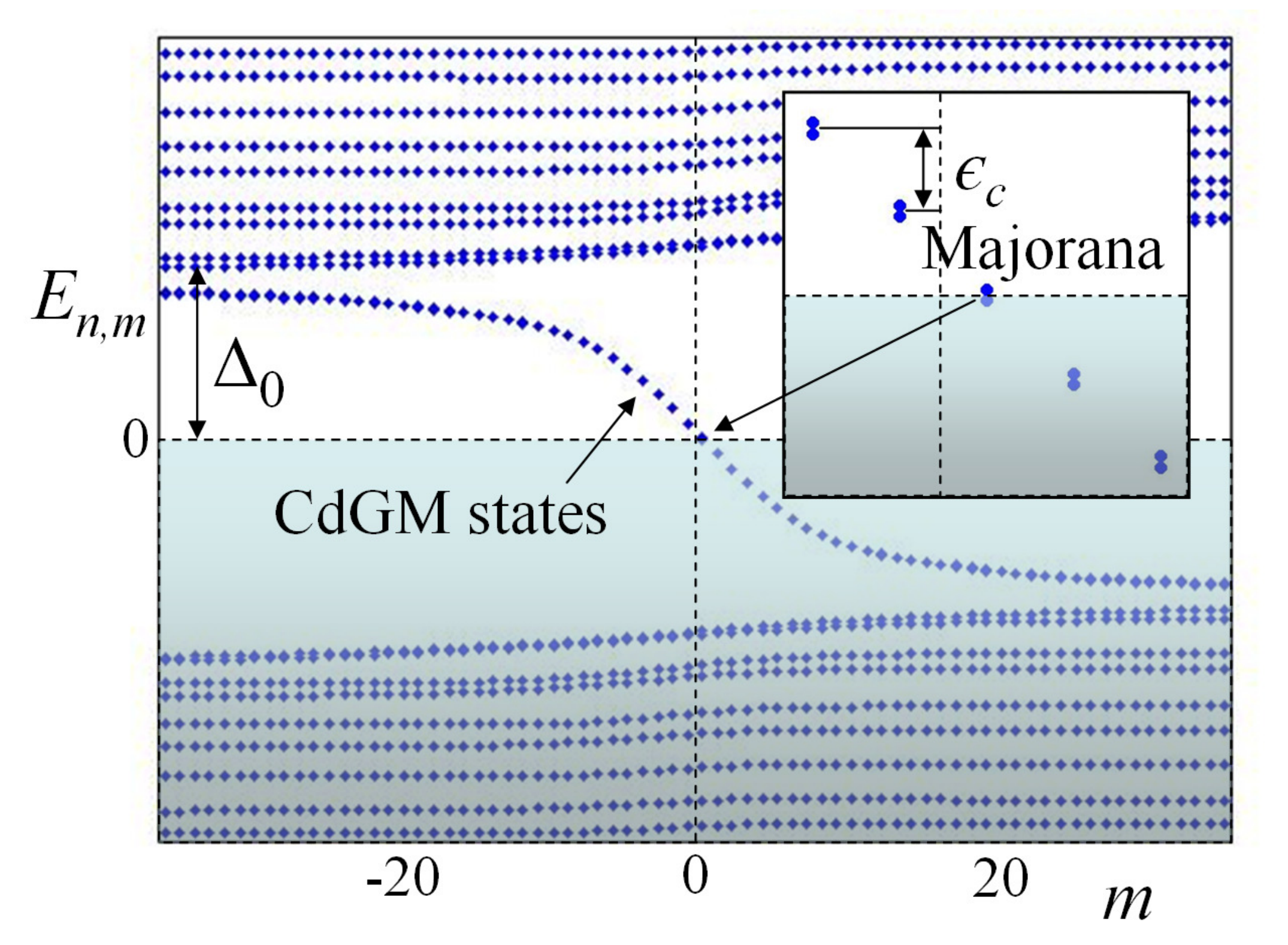}
\caption{ \label{Fig:Em} %
BdG spectrum $E_{n,m}$, of the vortex pair on the sphere, depicting
the CdGM core states. The inset shows that their double degeneracies
are split by weak tunneling between the poles. The positive energy
member of the doublet which saddles zero energy at $m = \half$, is
the Majorana state shared by the vortex and antivortex. }
\end{center}
\end{figure}

In Fig.~\ref{Fig:Em} we depict the BdG spectrum of the vortex pair
state as a function of $m$. The spectrum shows an expected symmetry
of the BdG equation, which implies that for every eigenvector $(u_n,
v_n)$ with energy $E_n$, the vector $(v_n^*,u_n^*)$ is also an
eigenvector with energy $-E_n$. Hence according to
Eqs.~(\ref{Eq:un})-(\ref{Eq:vn}),
\begin{eqnarray}  \label{Eq:unm_Enm}
    u_{n,m} & = & (v_{n',-m+1})^*, \\
    E_{n,m} & = & -E_{n',-m+1}. \nonumber
\end{eqnarray}

The continuum states above the gap $|E_n| > \Delta_0$ are extended,
while the branch that approaches zero is the \pxipy version of the
CdGM core states. Their dispersion is
\begin{eqnarray}  \label{Eq:Emc}
    E^c_m & \approx & (\half - m) \eC, \\
    \eC & = & \Do2eF, \nonumber
\end{eqnarray}
and their number is of order $\eF/\Delta_0$ \cite{KopninSalomaa}.

\begin{figure}[htb]
\begin{center}
\includegraphics[width=8cm,height=5.5cm,angle=0]{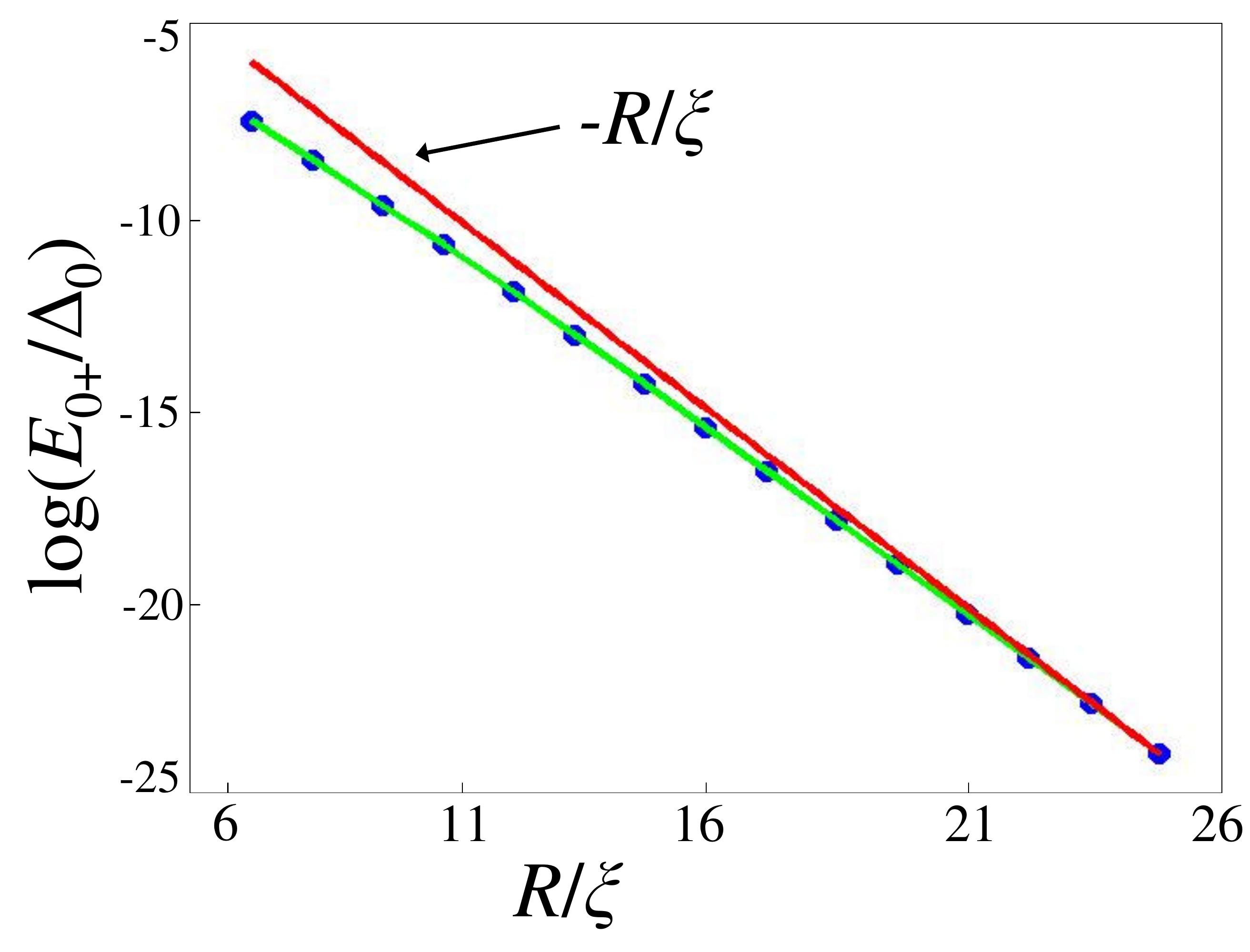}
\caption{ \label{Fig:E0} %
Energy of Majorana state $E_{0^+}$ as a function of sphere radius
$R$. The exponentially decreasing energy indicates tunnel splitting
between vortex and antivortex core states.}
\end{center}
\end{figure}

\begin{figure}[htb]
\begin{center}
\includegraphics[width=8cm,height=5.5cm,angle=0]{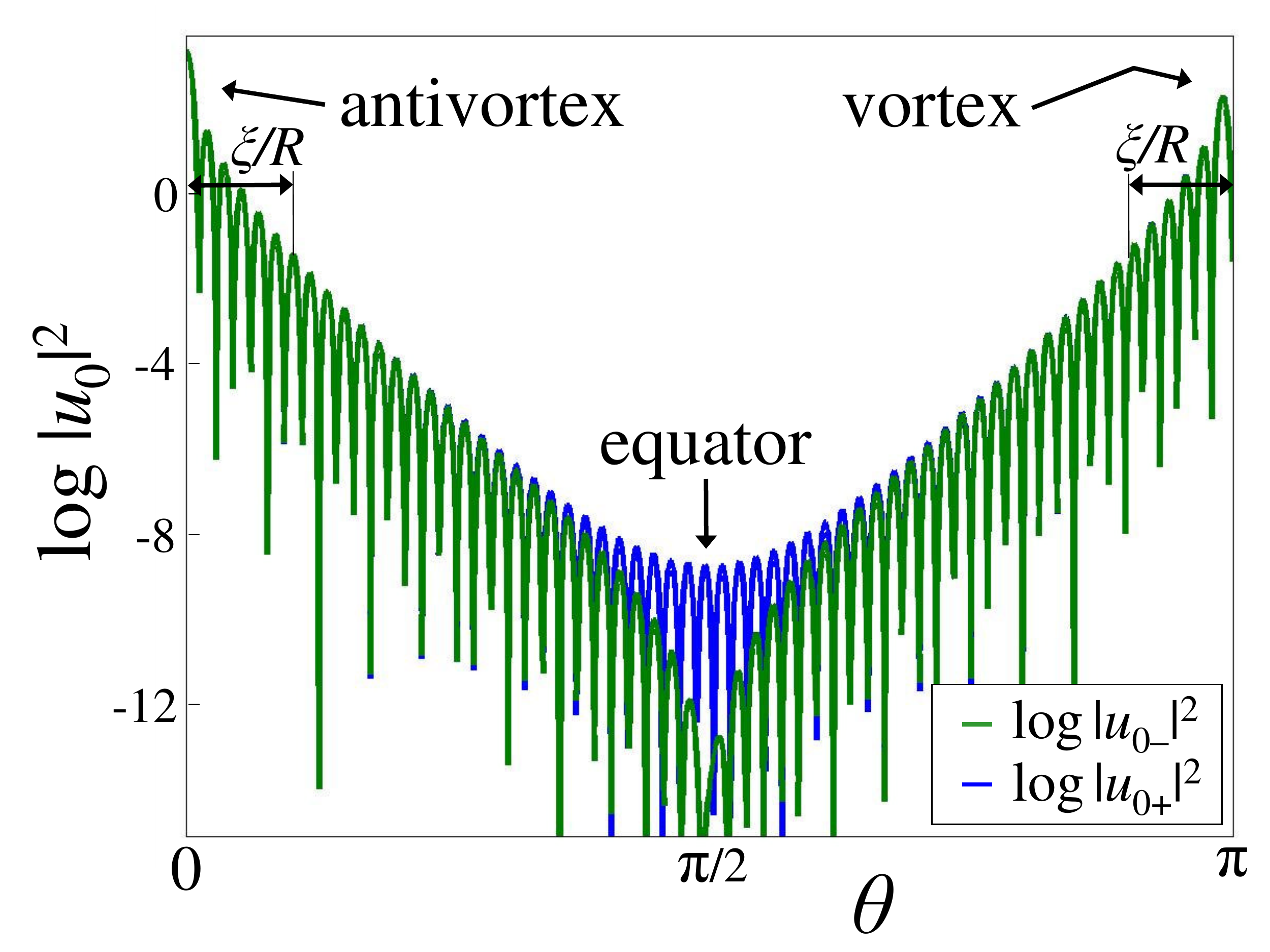}
\caption{ \label{Fig:u0v0} %
Probability densities of Majorana states $|u_0|^2$ versus latitude
on the sphere $\theta$. The smooth parts of the Majorana state
wavefunctions $u_{0^+}(\theta)$ and $u_{0^-}(\theta)$ are
approximately symmetric and antisymmetric with respect to reflection
about the equator $\theta=\pi/2$. Both $|u_{0^+}|^2$ (blue) and
$|u_{0^-}|^2$ (green) show the exponential localization in the
(anti)vortex cores.}
\end{center}
\end{figure}

As seen in the inset of Fig.~\ref{Fig:Em}, each CdGM state is almost
doubly degenerate. The splitting represents weak tunneling between
the north and south pole cores, and decreases exponentially with the
radius of the sphere $\delta E^c_m \sim \e^{-R/\xi}$ for $R \gg
\xi$. In particular, the lowest positive energy $E_{0^+}$, and its
negative companion $E_{0^-}$ (both at $m = \half$), approach zero as
$\e^{-R/\xi}$, is shown in Fig.~\ref{Fig:E0}. The probability
densities $|u_{0^+}(\theta)|^2 = |v_{0^-}(\theta)|^2$ and
$|u_{0^-}(\theta)|^2 = |v_{0^+}(\theta)|^2$ are depicted in
Fig.~\ref{Fig:u0v0}. The wavefunctions are symmetric and
anti-symmetric superpositions of the north and south localized core
states.

In the infinite sphere limit $E_{0^+} \approx E_{0^-} \approx 0$,
and the wavefunctions $u_0(\omgvec) \approx v_0(\omgvec)$ are
equally split between the north and south poles. The corresponding
BdG quasiparticle field operator near each pole $\eta(\omgvec) =
u_0(\omgvec) \psi(\omgvec) + v_0(\omgvec) \psi^\dag(\omgvec)$, is a
\emph{Majorana fermion} operator $\eta \approx \eta^\dag$.

\begin{figure}[htb]
\begin{center}
\includegraphics[width=8cm,height=10cm,angle=0]{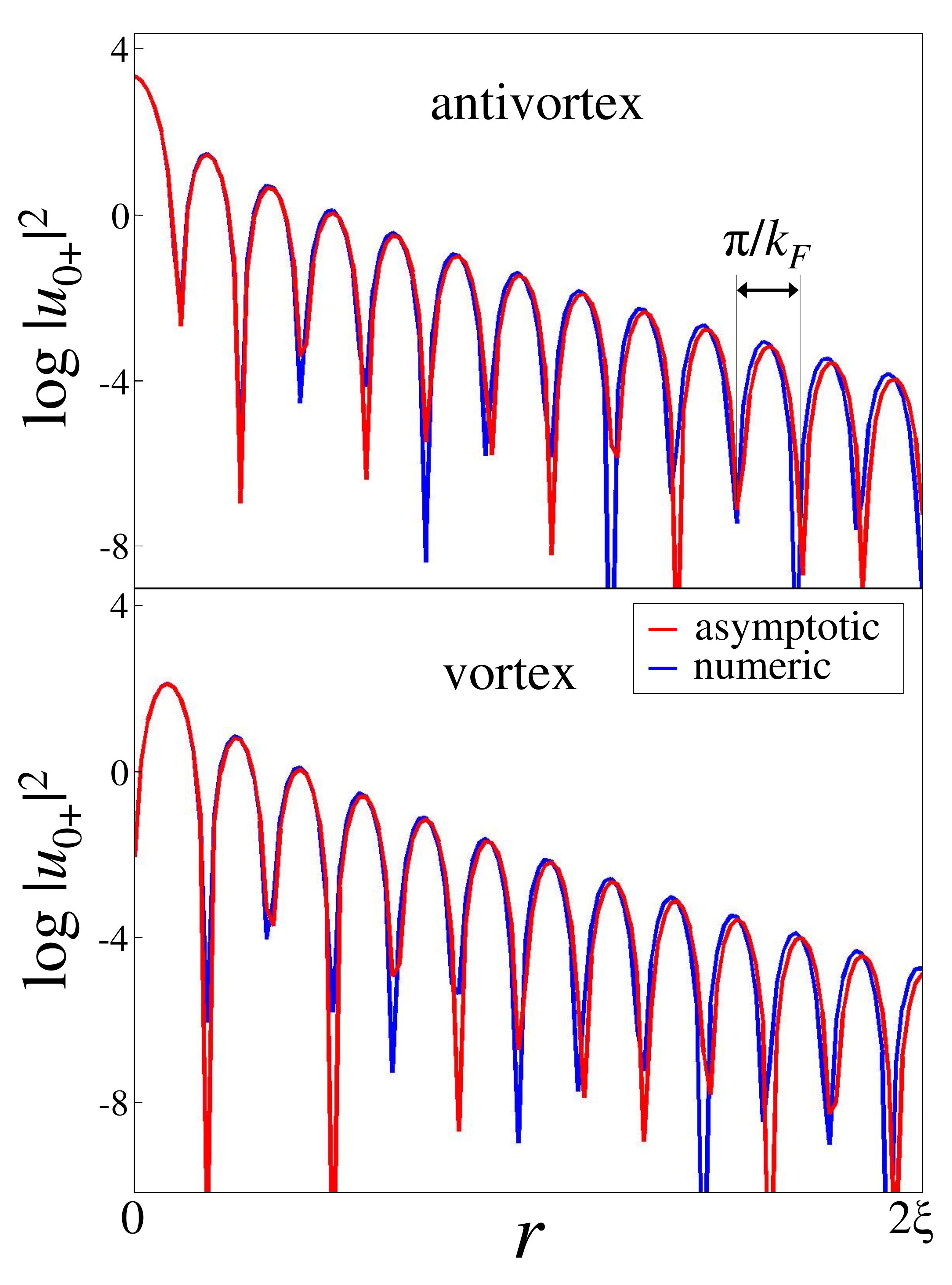}
\caption{ \label{Fig:u0_analytic} %
Probability distributions of Majorana state $|u_{0^+}|^2$ as a
function of the distance from the (anti)vortex center $r$, at the
cores vicinities. The wavefunctions have exponentially small weights
near the equator. The asymptotic (red) and the numeric (blue)
wavefunctions show excellent agreement.}
\end{center}
\end{figure}

The asymptotic behavior of the wavefunctions $u_0(r)$ in the plane
are \cite{GurarieRadzihosky, KopninSalomaa}
\begin{equation}   \label{Eq:u0}
    u_0(\xvec) \sim \left\{
        \begin{array}{ll}
            J_0(k_Fr) \e^{-{1 \over \pi} \int^r \mathrm{d}(r'/\xi) f_{\rm v}(r'/\xi)}
                            & \mbox{antivortex}\\
            J_1(k_Fr) \e^{-{1 \over \pi} \int^r \mathrm{d}(r'/\xi) f_{\rm v}(r'/\xi)}
                \e^{i\phi}  & \mbox{vortex}
        \end{array}  \right.
\end{equation}
where $f_{\rm v}$ was defined in Eq.~(\ref{Eq:Delta_pm}). We
confirmed this asymptotic behavior, as seen in
Fig.~\ref{Fig:u0_analytic}.

The physical reason behind the difference in Eq.~(\ref{Eq:u0})
between the vortex and the antivortex is that the phase winding of
the order parameter is determined by the sum of the vorticity and
the relative angular momentum. For a vortex the vorticity and the
angular momentum are aligned, which yields the phase winding
$\e^{2i\phi}$. In that case the condition for Majorana fermion
solution of the BdG equation, $u(\xvec) = v^*(\xvec)$, can be
fulfilled only by $u_0(\xvec) = \tilde{u}_0(r) \e^{i\phi}$, where
$\tilde{u}_0(r)$ vanishes at the origin (the $J_1(r)$ behavior is
expected, due to the azimuthal angular momentum 1). By contrast, for
an antivortex the order parameter is real, since the vorticity and
the relative angular momentum cancel each other. Thus the Majorana
fermion solution is real and radial $u_0(\xvec) = \tilde{u}_0(r)$,
and can be finite at the origin. (Here, $J_0(r)$ is expected, since
the angular momentum is zero). We will see in Sec.~\ref{Sec:LDOS}
that this difference is crucial for an experimental signature of the
Majorana state.

The excellent agreement between the asymptotic and numerical
wavefunctions, as seen in Fig.~\ref{Fig:u0_analytic}, reveals the
underlying physics of the core states. The radial profile of the
order parameter serves only as a confinement, and therefore
determines only the exponential decay of the wavefunction (and very
weakly the spacing between the core states). The short range part of
the wavefunction is controlled by a Bessel function, which is
determined only by the symmetry and the total angular momentum.
Furthermore, the specific pairing functions and pairing range do not
play any apparent role.


{\em \Swave.} In order to emphasize the unique behavior of the
chiral \mbox{{\it p}-wave} order parameter, we compare it to a
regular \mbox{\Swave} order parameter, characterized with the same
set of physical parameters.

An order parameter with \Swave symmetry in a plane, in the presence
of a vortex, will have the form
\begin{eqnarray}
    \Delta^S_{\pm}(\xvec,\xvec') & = & \Delta^S(\xvec-\xvec')
              f_{\rm v}(\bar{r}/\xi) e^{ \pm i\bar{\phi}} \label{Eq:DeltaS_pm}, \\
    \Delta^S (\xvec - \xvec')
        & = & \Delta_0 \frac{1}{4\pi\xi_p^{\phantom{p}2}}
              \e^{-\frac{(\xvec - \xvec')^2}{4\xi_p^{\phantom{p}2}}} \label{Eq:DeltaS},
\end{eqnarray}
where $f_{\rm v}$ is the same as in Eq.~(\ref{Eq:Delta_pm}).
$\Delta^S$ provides the \Swave pairing, with the same pairing range
$\xi_p$ of the \pxipy order parameter.

Implementing the order parameter on a sphere with a
vortex-antivortex pair at the poles is quite similar to the \pxipy
case:
\begin{eqnarray}  \label{Eq:DeltaS_vv}
    \Delta^S_V (\omgvec,\omgvec')
        & = & \Delta^S_p (\omgvec,\omgvec') F_V(\bar{\omgvec}),  \\
    \Delta^S_p (\omgvec,\omgvec')
        & = & \frac{\Delta_0} {4\pi\xi_p^{\phantom{p}2}}
              | \alpha \alpha'^* + \beta \beta'^* |^{ 2\Rxi },
\end{eqnarray}
where $F_V$ is unchanged.

Without the chirality, the order parameter can be expanded in the
regular spherical harmonics $Y_{lm}$ basis, with $l,m$ integers. In
this basis, the matrix forms of the kinetic term and the order
parameter are
\begin{eqnarray}
    T_{lm,l'm'} & = & \delta_{ll'} \delta_{mm'} \; \eF \left(
                      \frac{l(l + 1)} {l_F(l_F + 1)} + 1
                      \right) \label{Eq:TSlmlm}, \\
    \Delta^S_{lm,l'm'} & = & -\delta_{m', 1-m} \Delta_0 \sqrt{ {\textstyle {1 \over 16\pi} }
                           (2l + 1) (2l' + 1) }  \label{Eq:DSlmlm} \\
                     &   & \times \left( D^S_l + D^S_{l'} \right)
                            \sum_{L} \sqrt{ 2L + 1 } \; f^V_L  \nonumber\\
                     &    & \qquad \times \Wj{l}{l'}{L}{0}{0}{0}
                                          \Wj{l}{l'}{L}{-m}{m-1}{1},  \nonumber\\
    D^S_l & \approx & \e^{ -l^2 \xiR }.
\end{eqnarray}
Our method of computing $\Delta^S_{lm,l'm'}$ is almost the same as
that for $\Delta_{lm,l'm'}$ (Appendix \ref{App:DVlmlm}), but with
$q$ set to $0$ and integer $l,m$. $D^S_l$ is given essentially by
Eq.~(\ref{Eq:aa_bb_3}), and is approximated in a way analogous to
$D_l$ (Eq.~\ref{Eq:Dl}), but without the chirality factor $l/l_F$.

\begin{figure}[htb]
\begin{center}
\includegraphics[width=8cm,height=5cm,angle=0]{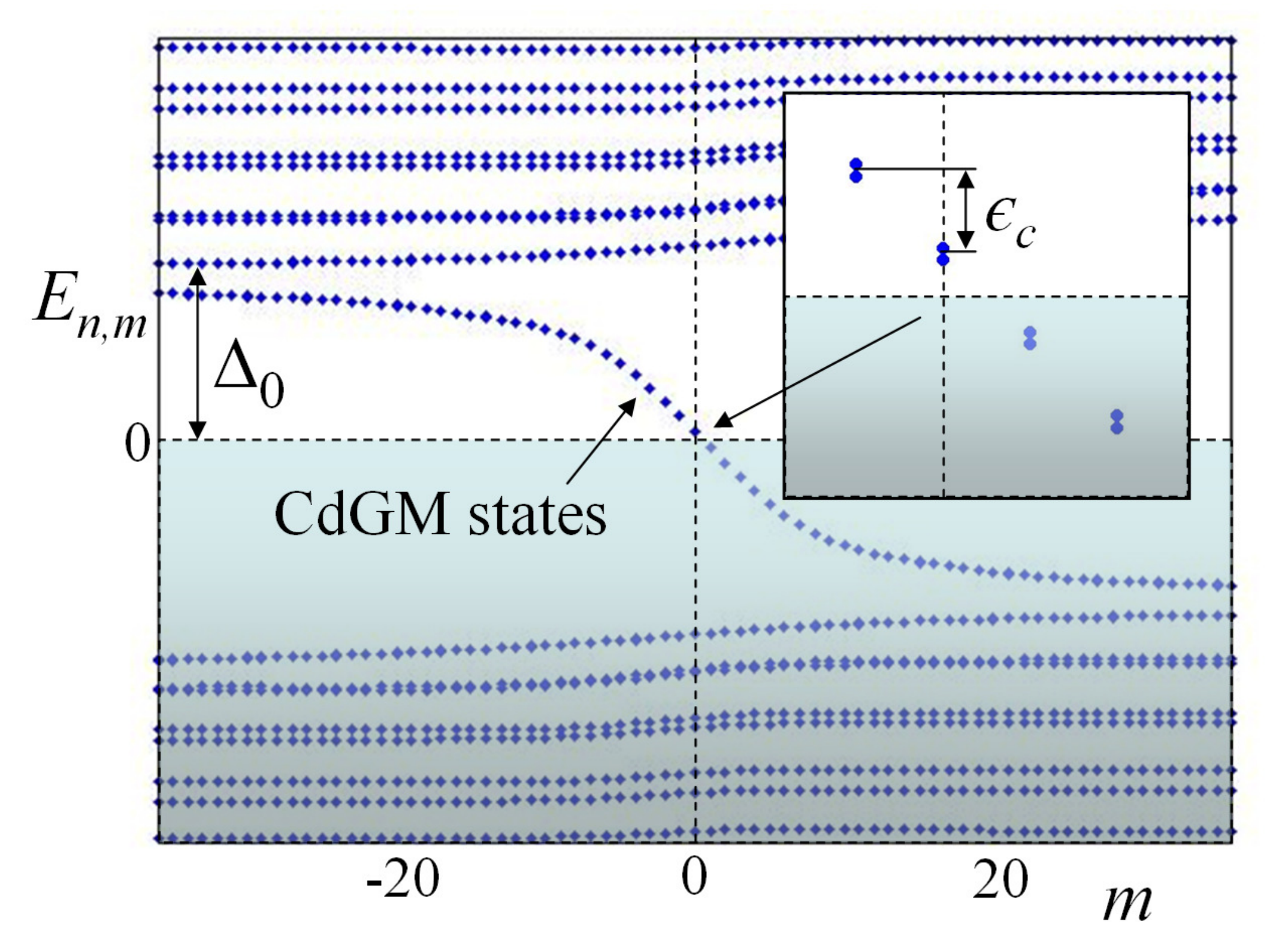}
\caption{ \label{Fig:Em_S} %
The BdG spectrum $E_{n,m}$, of an \Swave superconductor, with the
same physical parameters of the \pxipy in Fig.~\ref{Fig:Em}. The
distinction from the \pxipy case can be noticed in the inset, where
the energies of the CdGM states are shifted by half the level
spacing, therefore lacking a Majorana state which saddles zero
energy.}
\end{center}
\end{figure}

The BdG spectrum of the \Swave system is generally similar to that
of the \pxipy system, as seen in Fig.~\ref{Fig:Em_S}, which was
plotted with the same parameters that were used to plot
Fig.~\ref{Fig:Em}. Both the extended states, with $|E_n| >
\Delta_0$, and the CdGM states, with $E^c_m \approx (\half - m)
\eC$, are present. Here $m$ is an integer, the energies of the CdGM
states are shifted by $\half\eC$ compared to the \pxipy case, and
the Majorana state is absent.

Far from the vortex core, the \Swave and the \pxipy show almost the
same gapped spectrum. Fourier transforming the BdG equation in the
plane (Eqs.~\ref{Eq:BdG}, \ref{Eq:Delta} and \ref{Eq:DeltaS}, see
Eq.~\ref{Eq:Dk}), and diagonalizing the BdG matrix, gives
\begin{eqnarray} \label{Eq:Ek}
    E^{p_x\!+\!ip_y}_{\boldsymbol{k}}
        & = & \sqrt{ ( k^2 / 2m^* - \eF )^2 +
                     ( \Delta_0 {k \over k_F} \e^{-k^2 \xi_p^{\phantom{p}2}} )^2 }, \\
    E^S_{\boldsymbol{k}} & = & \sqrt{ ( k^2 / 2m^* - \eF )^2 +
                     ( \Delta_0 \e^{-k^2 \xi_p^{\phantom{p}2}} )^2 },
\end{eqnarray}
where $k = ( k_x^2 + k_y^2 )^{1/2}$.

The small difference between the two cannot be distinguished in a
tunneling experiment. Nevertheless, the disparity in the CdGM
states, and especially the existence of the Majorana state, may be
observed by taking a difference in the tunneling LDOS, as we will
see in Sec.~\ref{Sec:LDOS}.

The last point to be mentioned is the insensitivity of the above
results to the choice of hamiltonian parameters. According to
Eqs.~(\ref{Eq:Tlmlm})-(\ref{Eq:Dl}) there are three free
dimensionless physical parameters: $\Delta_0/\eF$, $l_F$ and
$\xi_p/R$, while $\xi/R$ is determined by \mbox{$l_F \cdot
\Delta_0/\eF$} (according to Eq.~\ref{Eq:xi}). However, the
asymptotic wavefunction (Eq.~\ref{Eq:u0}) implies that the
exponential envelope is controlled by $\xi$. Moreover, the energy of
the Majorana state is also determined by $\xi$. We confirmed this
numerically, for the physical regime $\Delta_0 \ll \eF$. $k_F$,
which is approximately $l_FR$, controls the oscillation frequency of
the wavefunction, as seen in Eq.~\ref{Eq:u0}. The gap $\Delta_0$ and
$\eF$ determine the energy spacing of the core states $\eC$ and
their number. $\xi_p$ has negligible effect, up to $k_F\cdot\xi_p
\leq 2\pi$, meaning that the pairing range has no significant effect
either on the wavefunctions or on the energies of the core states.
Additionally, modifying the vortex profile of Eq.~(\ref{Eq:F_V}) --
to, for example, $\tanh(\sin\theta \cdot R/\xi)$ -- only modifies
the wavefunctions slightly inside the cores, according to
Eq.~\ref{Eq:u0}.


\section{Disorder}
\label{Sec:Disorder} %
Disorder mixes states with different angular momentum $m$, thus
making the BdG equation extremely hard to solve analytically. We
study numerically the effect of disorder on the Majorana state, by
the addition of a white noise random real potential to the BdG
equation (Eq.~\ref{Eq:BdG}), given by
\begin{equation}  \label{Eq:whitenoise}
    W(\omgvec) = \sum_{l,m}^{\lL} w_{lm} Y_{lm}(\omgvec),
\end{equation}
where the $w_{lm}$'s are independently identically distributed, with
$w_{l-m} = (-1)^m w_{lm}^*$. $\lL$ is an ultra violet cut-off. We
want the potential to be independent of the radius of the sphere via
$\lL$, therefore we take the real and imaginary parts of $w_{lm}$ to
be uniformly distributed in the interval $\left[ -\frac{\sqrt{ 6\pi
}\w}{\lL}, \frac{\sqrt{ 6\pi }\w}{\lL} \right]$, where $\w$ is given
in units of energy. This gives
\begin{equation}  \label{Eq:W2}
    \langle W^2(\omgvec) \rangle = \ww,
\end{equation}
as shown in Appendix \ref{App:Wnoise} and demonstrated in
Fig.~\ref{Fig:noise}.

\begin{figure}[htb]
\begin{center}
\includegraphics[width=7cm,height=13cm,angle=0]{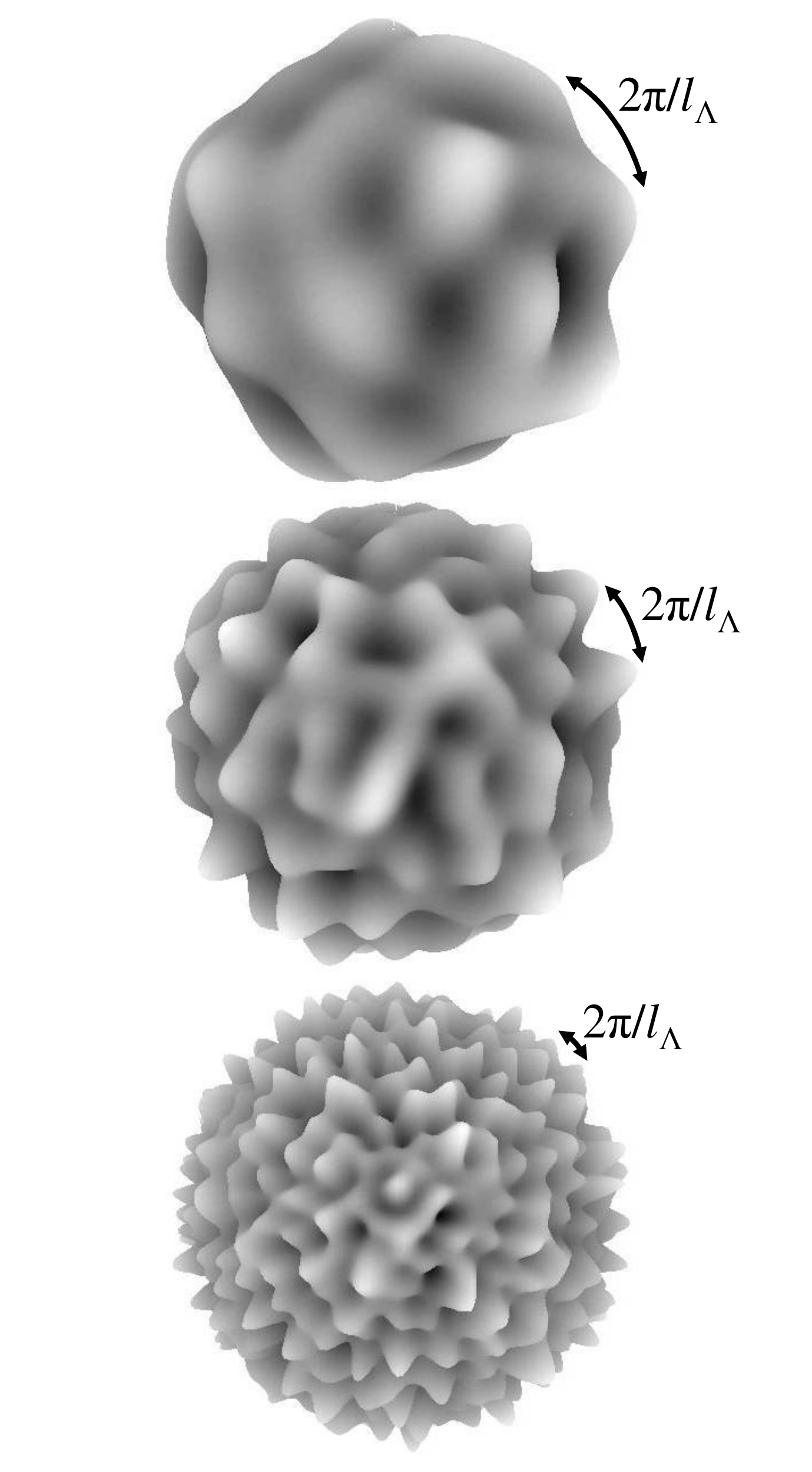}
\caption{ \label{Fig:noise} %
Three white noise potentials $W(\omgvec)$ on the sphere, taken from
the same distribution of harmonics components $w_{lm}$, but of
increasing high angular momentum cut-off $\lL$ (from top to bottom).
Each time $\lL$ is multiplied by 2, while the disorder strength $\w$
remains constant. }
\end{center}
\end{figure}

The disorder breaks the azimuthal symmetry, so that $m$ is no longer
a good quantum number. The matrix elements of $W(\omgvec)$ are
\begin{eqnarray} \label{Eq:Wlmlm}
    W_{lm,l'm'} & = & (-1)^{m-\half} \sqrt{ {\textstyle {1 \over 4\pi} }
                      (2l + 1) (2l' + 1) } \nonumber \\
                &   & \times \sum_{L}^{\lL} \sqrt{ 2L + 1 } \; w_{L,m'-m} \\
                &   & \qquad \times \Wj{l}{l'}{L}{\half}{-\half}{0}
                                    \Wj{l}{l'}{L}{m}{-m'}{m'-m} \nonumber,
\end{eqnarray}
as derived in Appendix \ref{App:Wnoise}.

Fig.~\ref{Fig:E_noise} depicts the disorder averaged energy of the
Majorana state $E_{0^+}$ versus $R$ for increasing $\w$. It can be
seen that both the average energy (solid) and the standard deviation
(error bars) decay exponentially in the regime $\w < \eF$. Thus we
conclude that the exponential drop of the Majorana state energy with
increasing system size survives moderate disorder.

\begin{figure}[htb]
\vspace{0cm}
\begin{center}
\includegraphics[width=8cm,height=13cm,angle=0]{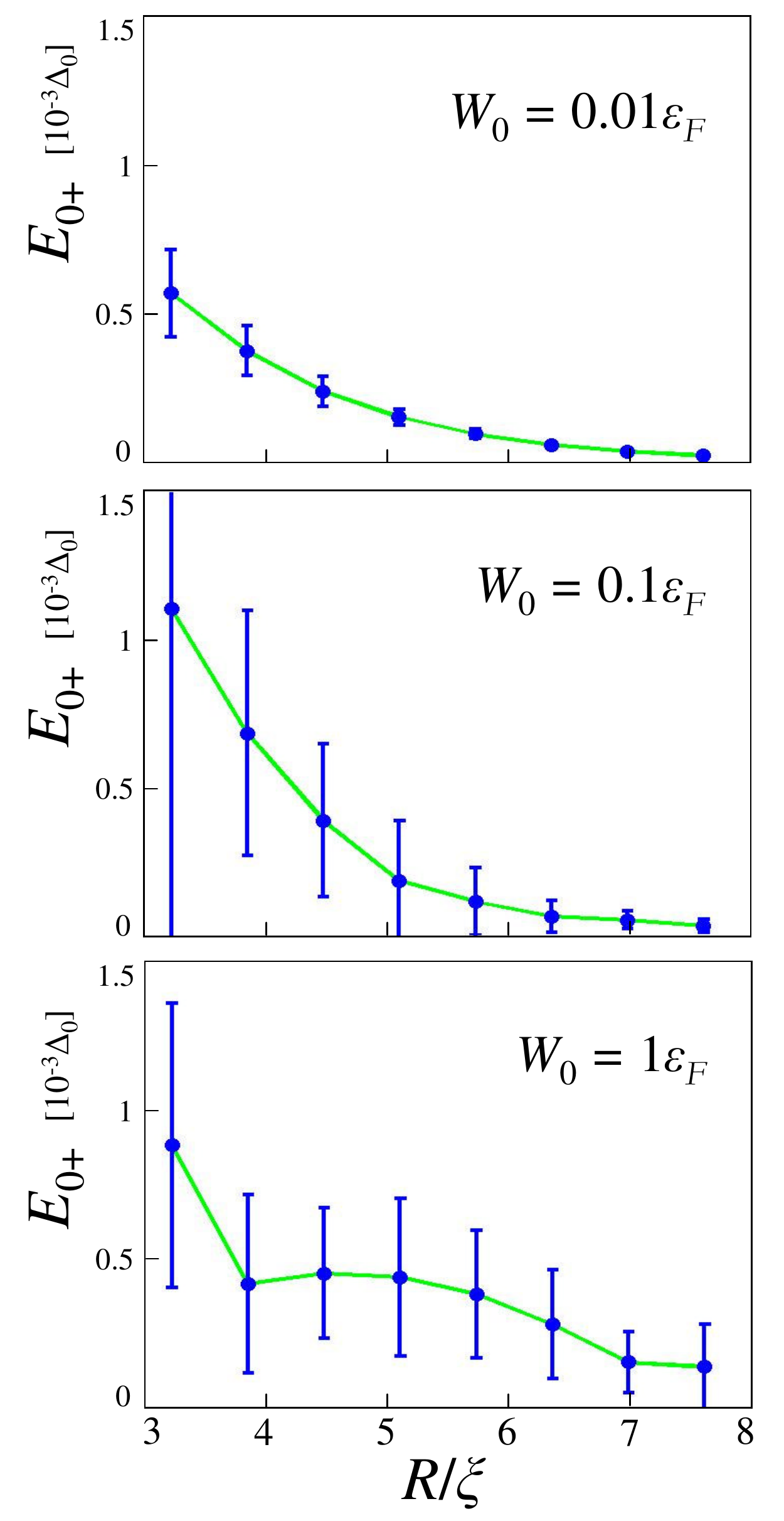}
\caption{ \label{Fig:E_noise} %
Disorder averaged energy of Majorana state $E_{0^+}$ versus $R/\xi$
for increasing disorder strength $\w$ (from top to bottom). The
exponential decay survives in the regime $\w < \eF$. }
\end{center}
\end{figure}


\section{System with an edge}
\label{Sec:Edge} %
The \pxipy state has broken time-reversal symmetry, implying there
are chiral modes which are exponentially localized at the edge of
the sample. For a disk of radius $R$, the energies of the edge
states are expected to have the form $E_m^{edge} \propto m / R$,
where the angular momentum $m$ is a half integer, due to the
anti-periodic boundary condition of the spin polarized fermions
\cite{ReadGreen,Fendley}.

In the presence of a half-quantum vortex, the boundary condition on
the BdG wavefunctions is periodic, and $m$ is an integer, with the
Majorana state having the $m = 0$ quantum number. A fermion
occupation number state is created from a combination of Majorana
state on the edge and in the vortex core. This was recently
investigated numerically \cite{Mizushima}.

In order to create an edge at latitude $\tE$, we add a strong
potential of the form
\begin{eqnarray}  \label{Eq:WE}
    W_E(\theta) & = & \frac{2\eF} { \e^{ (\tE - \theta) 2R/\xi } + 1 } \nonumber \\
                & = & \sum_L w^E_L Y_{L0}(\omgvec),
\end{eqnarray}
which defines $w^E_L$. Note that at the edge $W_E(\tE) = \eF$, which
sets the density to zero. The width of the potential was chosen to
be the longer length scale $\xi$.

For a uniform superconductor with an edge, the order parameter is of
the form
\begin{eqnarray}
    \Delta_E(\omgvec,\omgvec') & = &
            \Delta_p(\omgvec,\omgvec') F_E(\bar{\omgvec}), \label{Eq:DE}\\
    F_E(\theta) & = & \left\{ \begin{array}{cc}
                              \tanh \left( (\tE - \theta) R/\xi \right)
                                & 0   \leq \theta \leq \tE \\
                              0 & \tE \leq \theta \leq \pi
                                 \end{array} \right.     \nonumber \\
             & = & \sum_L f^E_L Y_{L0}(\omgvec),   \label{Eq:fE}
\end{eqnarray}
where the pairing $\Delta_p(\omgvec,\omgvec')$ is defined in
Eq.~(\ref{Eq:Delta_p}), and the envelope $F_E(\bar{\omgvec})$ equals
to zero for $\tE \leq \theta \leq \pi$ due to self consistency. The
matrix elements of $W_E$ and $\Delta_E$ appear in Appendix
\ref{App:MatrixE}, which also shows that $m$ is a good quantum
number.

\begin{figure}[htb]
\begin{center}
\vspace{0cm}
\includegraphics[width=8cm,height=5cm,angle=0]{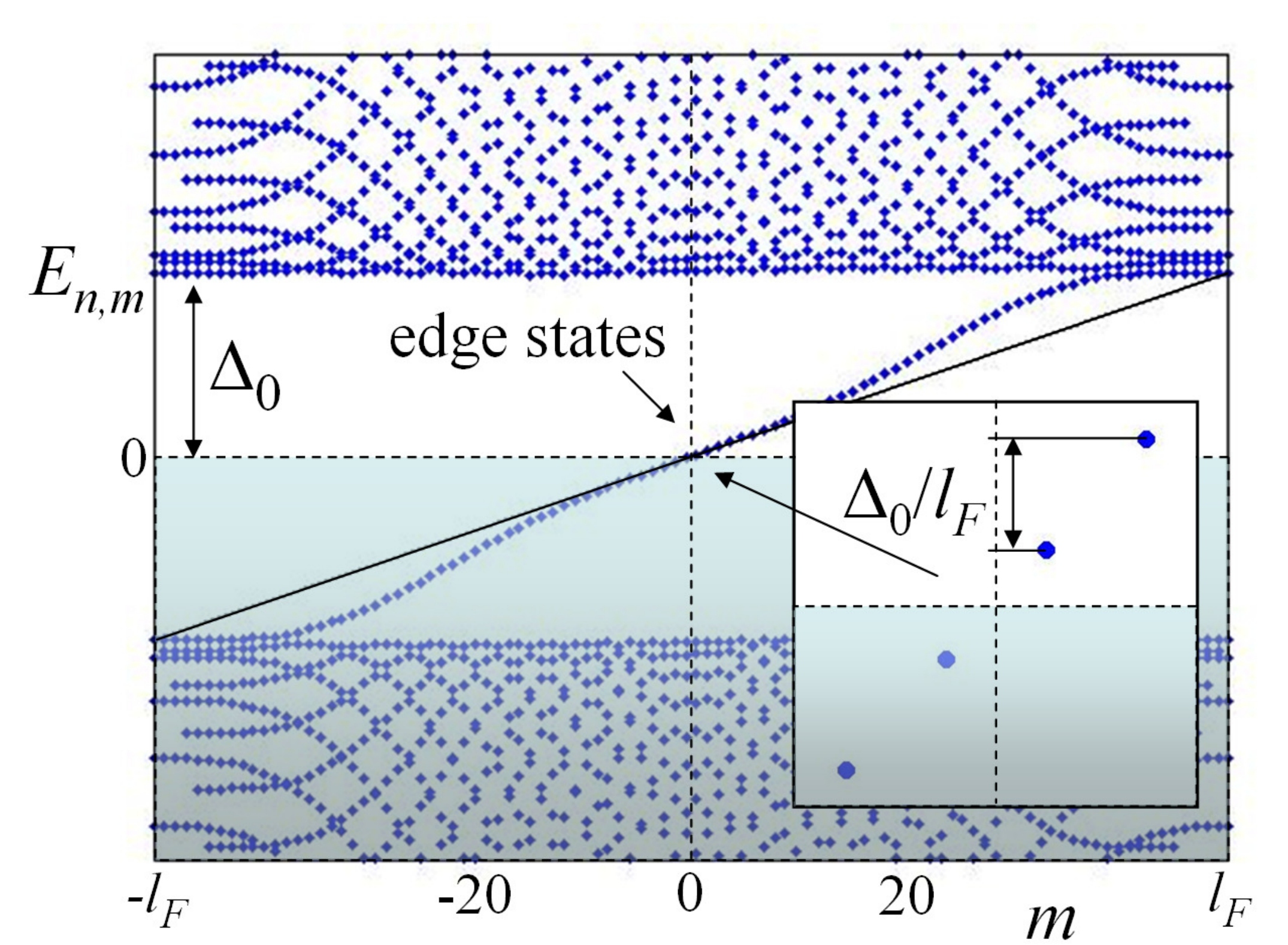}
\vspace{0cm} \caption{ \label{Fig:Em_Edge} %
BdG spectrum $E_{n,m}$, of an edge without a vortex, depicting edge
states. The solid black line is $m\Delta_0/l_F$. Here $R/\xi = 7.6$,
while the fit improves for larger $R/\xi$.}
\end{center}
\end{figure}

Fig.~\ref{Fig:Em_Edge} depicts the BdG spectrum of an edge without a
vortex. The states above the gap $|E_n| > \Delta_0$ are extended,
while the branch is composed of the chiral edge state. However,
since there is only a single edge, these states are not degenerate.

\begin{figure}[htb]
\begin{center}
\vspace{0cm}
\includegraphics[width=8cm,height=5.5cm,angle=0]{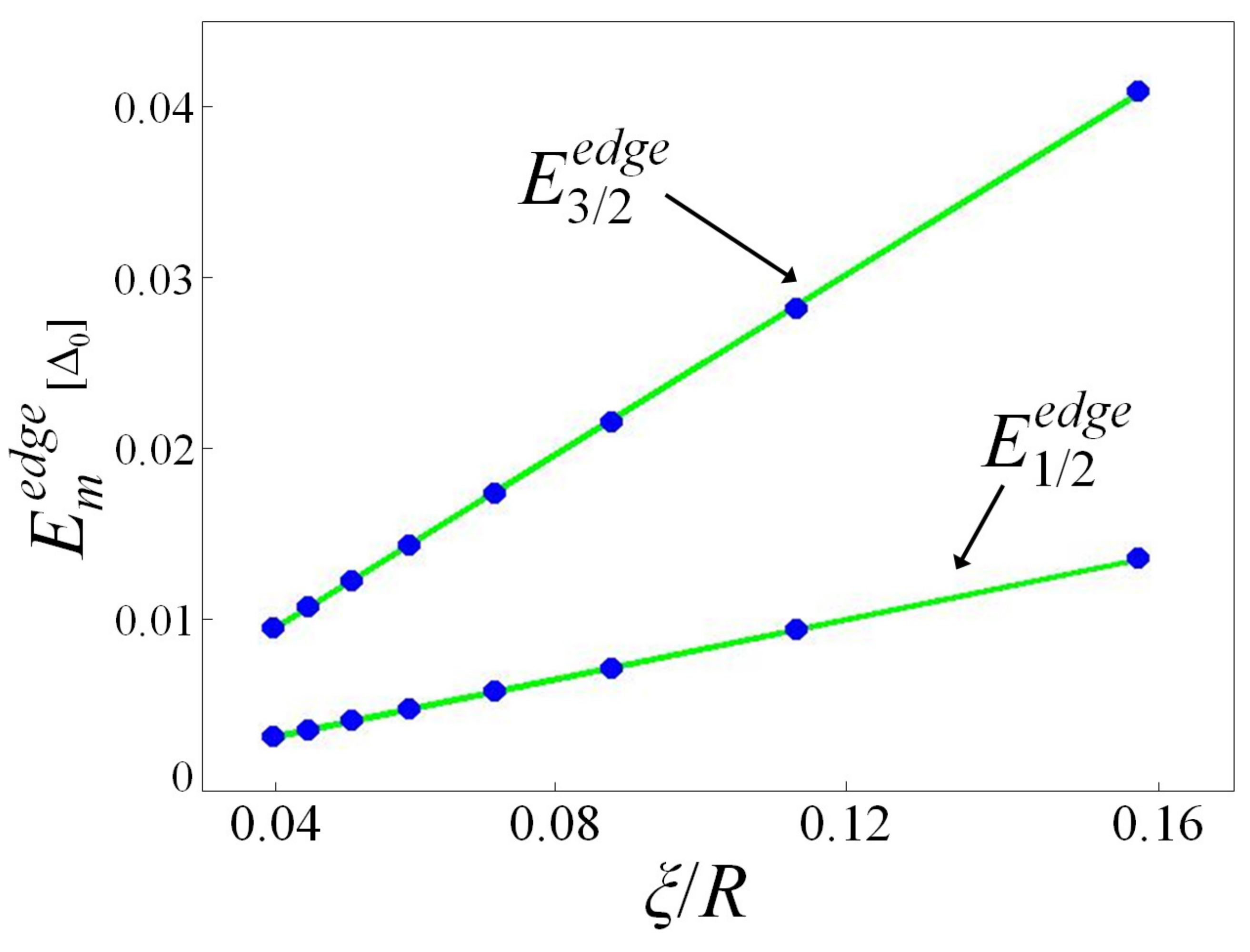}
\vspace{0cm} \caption{ \label{Fig:E_Edge} %
Energies of the two lowest edge states without a vortex
$E_m^{edge}$, for $m = \half$ and $m = {3 \over 2}$, versus $\xi/R$,
show the $1/R$ decay.}
\end{center}
\end{figure}

In a half-infinite plane the dispersion of the edge states is given
by \cite{Fendley} $E_k^{edge} \approx \Delta_0 {k \over k_F}$, with
the velocity $v_{edge} = {\Delta_0 \over k_F}$. Conversion to
spherical geometry with $R \gg \xi$, gives
\begin{equation}  \label{Eq:Em_E}
    E_m^{edge} \approx \Delta_0 {m \over l_F}.
\end{equation}
The black line in Fig.~\ref{Fig:Em_Edge} depicts this approximation
for $R/\xi = 7.6$. The approximation improves for larger values of
$R/\xi$. Moreover, Fig.~\ref{Fig:E_Edge} shows that $E_m^{edge}
\propto 1/R$, as expected.

When considering the antivortex in the north pole, the potential
remains the same, while the order parameter becomes
\begin{eqnarray}
    \Delta_{V\!E}(\omgvec,\omgvec') & = &
            \Delta_p(\omgvec,\omgvec') F_{V\!E}(\bar{\omgvec}), \label{Eq:DVE}\\
    F_{V\!E}(\theta) & = & \e^{ i\phi} \left\{ \begin{array}{cc}
                              \tanh \left(
                              \sin \left( \pi {\theta \over \tE} \right) {R \over \xi} \right)
                                & 0   \leq \theta \leq \tE \\
                              0 & \tE \leq \theta \leq \pi
                                 \end{array} \right.     \nonumber \\
             & = & \sum_L f^{V\!E}_L Y_{L1}(\omgvec).   \label{Eq:fVE}
\end{eqnarray}
Note that the vortex envelope $F_{V\!E}$ is no longer symmetric with
respect to the equator, and so includes $f^{V\!E}_L \neq 0$ for even
$L$'s, unlike $F_V$ (Eq.~\ref{Eq:fVL}). The matrix elements of
$\Delta_{V\!E}$ appear in Appendix \ref{App:MatrixE}.

\begin{figure}[htb]
\begin{center}
\vspace{0cm}
\includegraphics[width=8cm,height=5cm,angle=0]{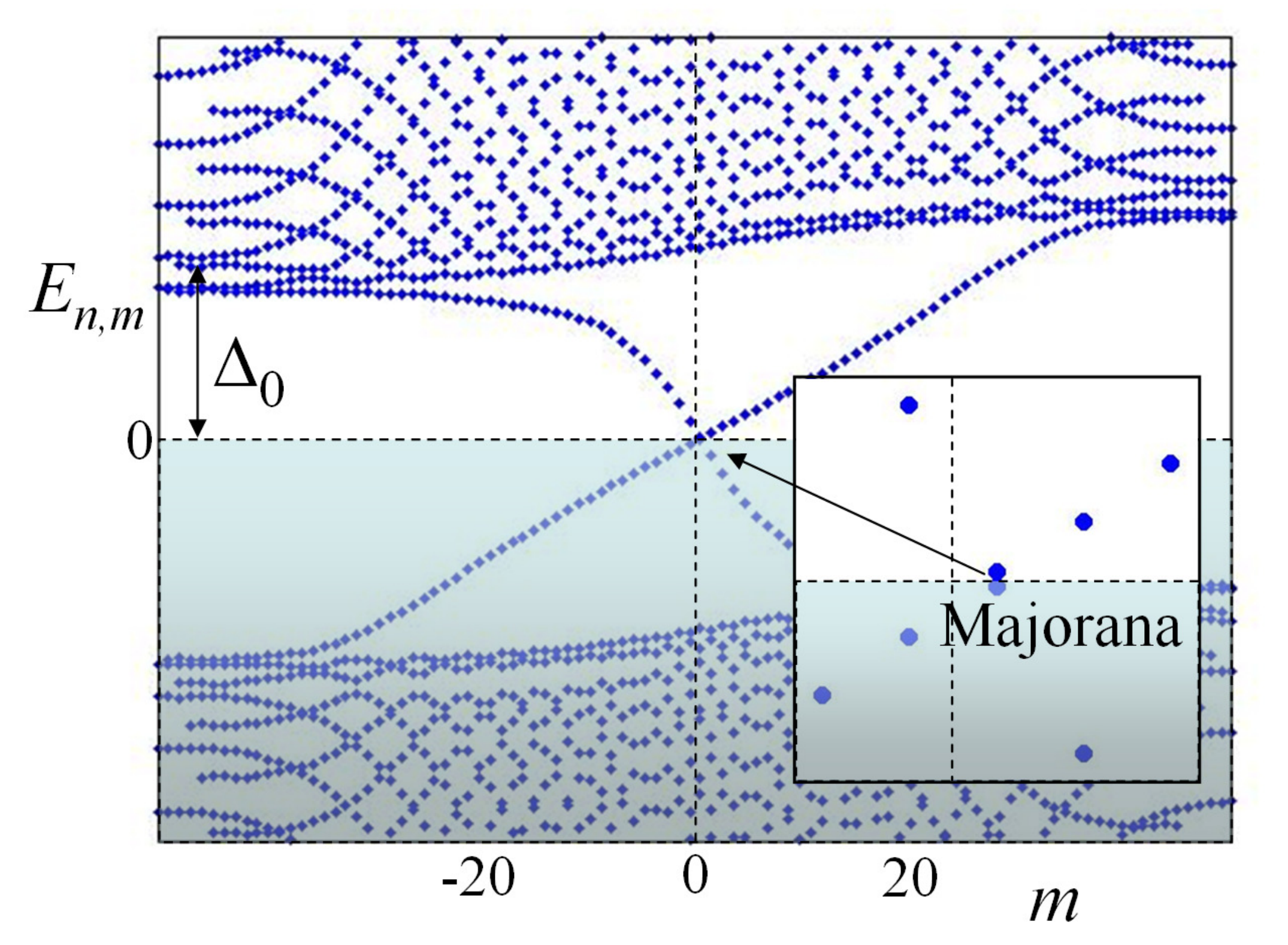}
\vspace{0cm} \caption{ \label{Fig:Em_EdgeV} %
BdG spectrum $E_{n,m}$, of an antivortex with an edge, showing both
the edge states and CdGM states. The Majorana state is present, with
an exponentially small energy, as seen in the inset.}
\end{center}
\end{figure}

The resultant BdG spectrum, $E_{n,m}$, is shown in
Fig.~\ref{Fig:Em_EdgeV}. The edge states and CdGM states appear in
two separate branches. The only quasi degenerate state is the
Majorana state at $m = \half$.

\begin{figure}[htb]
\begin{center}
\vspace{0cm}
\includegraphics[width=8.5cm,height=12cm,angle=0]{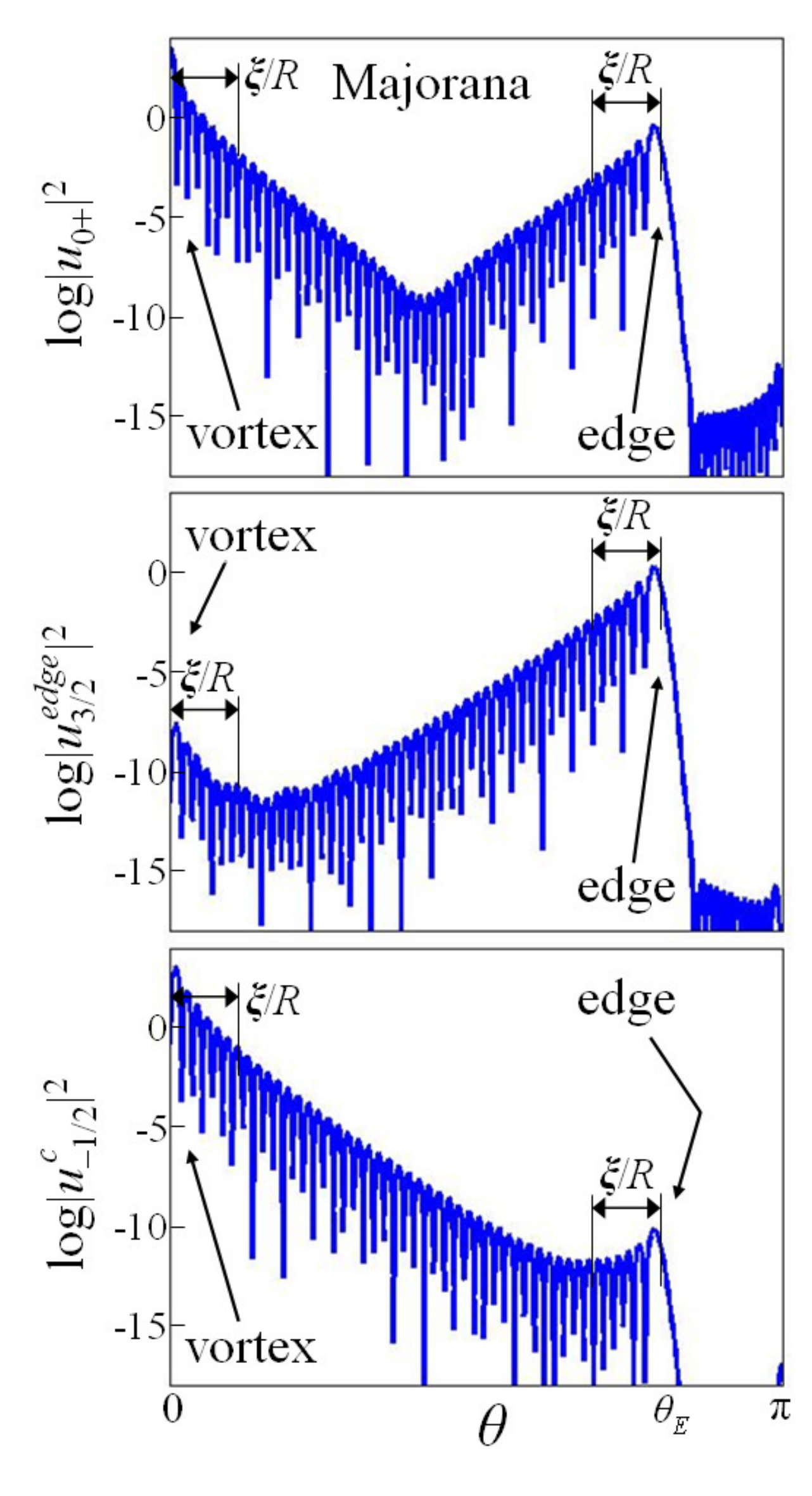}
\vspace{-0.3cm} \caption{ \label{Fig:wf_EdgeV} %
$\log|u(\theta)|^2$ of the Majorana state $u_{0^+}$ (top), the first
excited edge state $u_{3/2}^{edge}$ (middle), and the first excited
CdGM state $u_{-1/2}^c$ (bottom). The Majorana state is split
between the edge and the vortex core, while in each branch of
excitation the wavefunction is concentrated \emph{either} in edge
\emph{or} in the vortex core.}
\end{center}
\end{figure}

\begin{figure}[htb]
\begin{center}
\vspace{0cm}
\includegraphics[width=8.2cm,height=10cm,angle=0]{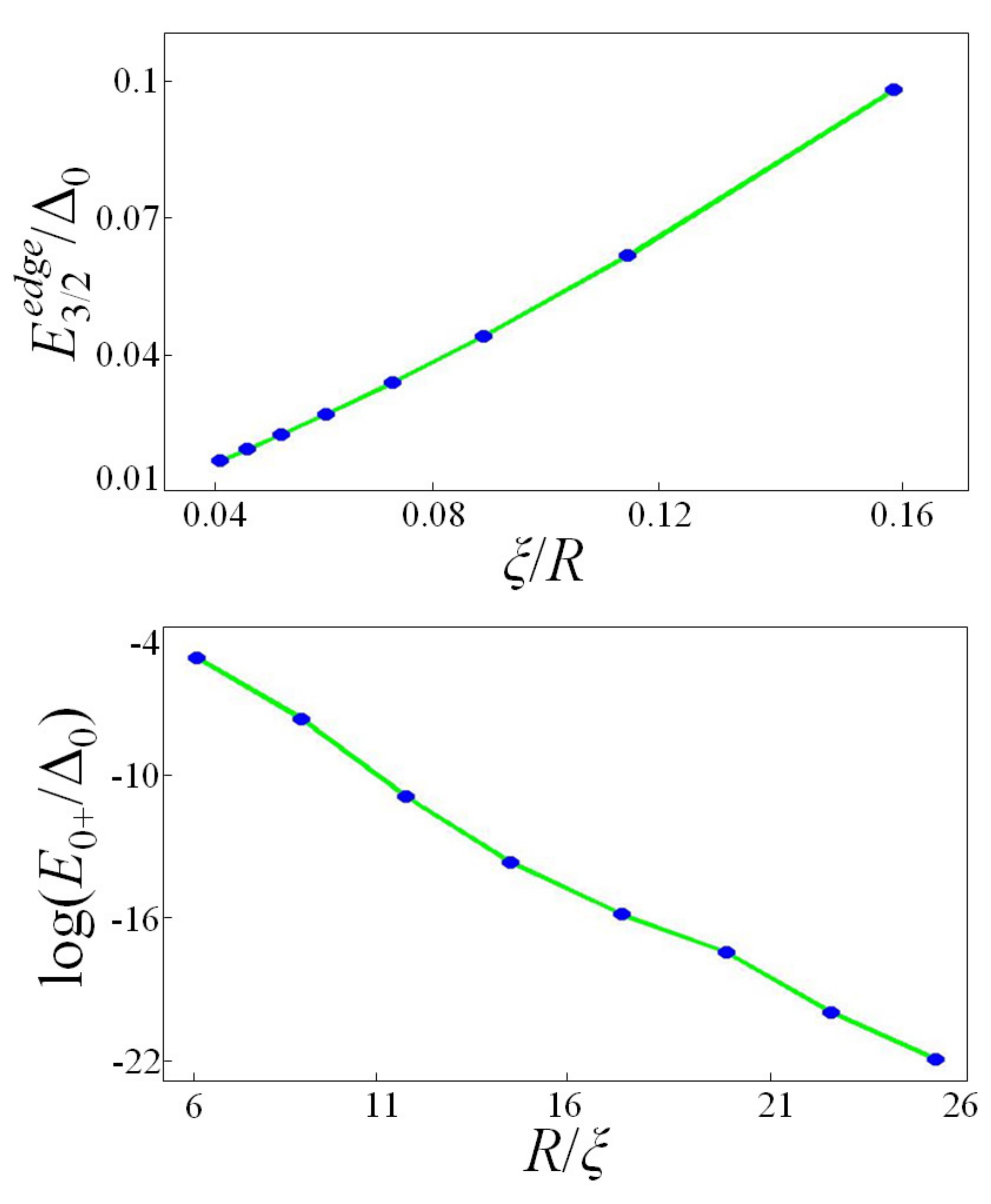}
\vspace{-0.3cm} \caption{ \label{Fig:E_EdgeV} %
Energies of Majorana state $E_{0^+}$ (bottom) and first excited edge
state $E_{3/2}^{edge}$ (top) of an antivortex with an edge, as a
function of the radius of sphere $R$. The energy of the Majorana
state decays exponentially, while the energy of the edge state
scales as $1/R$.}
\end{center}
\end{figure}

Fig.~\ref{Fig:wf_EdgeV} depicts the wavefunctions of the Majorana
state $u_{0^+}(\theta)$, the first excited edge state
$u_{3/2}^{edge}(\theta)$ at $m = {3 \over 2}$, and the first excited
CdGM state $u_{-1/2}^c(\theta)$ at $m = -\half$. It can be seen that
the Majorana state has almost equal support on the edge and in the
vortex core (and is exponentially localized in both), while the edge
state and the CdGM state are concentrated {\em either} at the edge
{\em or} at the vortex core, respectively. Fig.~\ref{Fig:E_EdgeV}
confirms that as expected, the energy of the Majorana state
$E_{0^+}$ decays to zero as $\e^{-R/\xi}$, while the energy of the
first excited edge state $E_{3/2}^{edge}$ scales as $1/R$.

In summary, the spherical geometry enables an easy visualization of
edge effects, by the addition of a confining potential. The expected
edge states appear in any case, while the Majorana state appears
only in the presence of a vortex.


\section{Tunneling local density of states}
\label{Sec:LDOS} %
The energy gap and the coherence length of a superconductor can be
detected by tunneling of electrons to its surface. For high spatial
resolution, the tunneling also detects the low energy excitation
spectrum inside a vortex core. In this section we show that
tunneling experiment also provides direct signatures of the symmetry
of the order parameter, and the existence of the Majorana state.

\begin{figure}[htb] \vspace{-0.3cm}
\begin{center}
\includegraphics[width=7cm,height=13cm,angle=0]{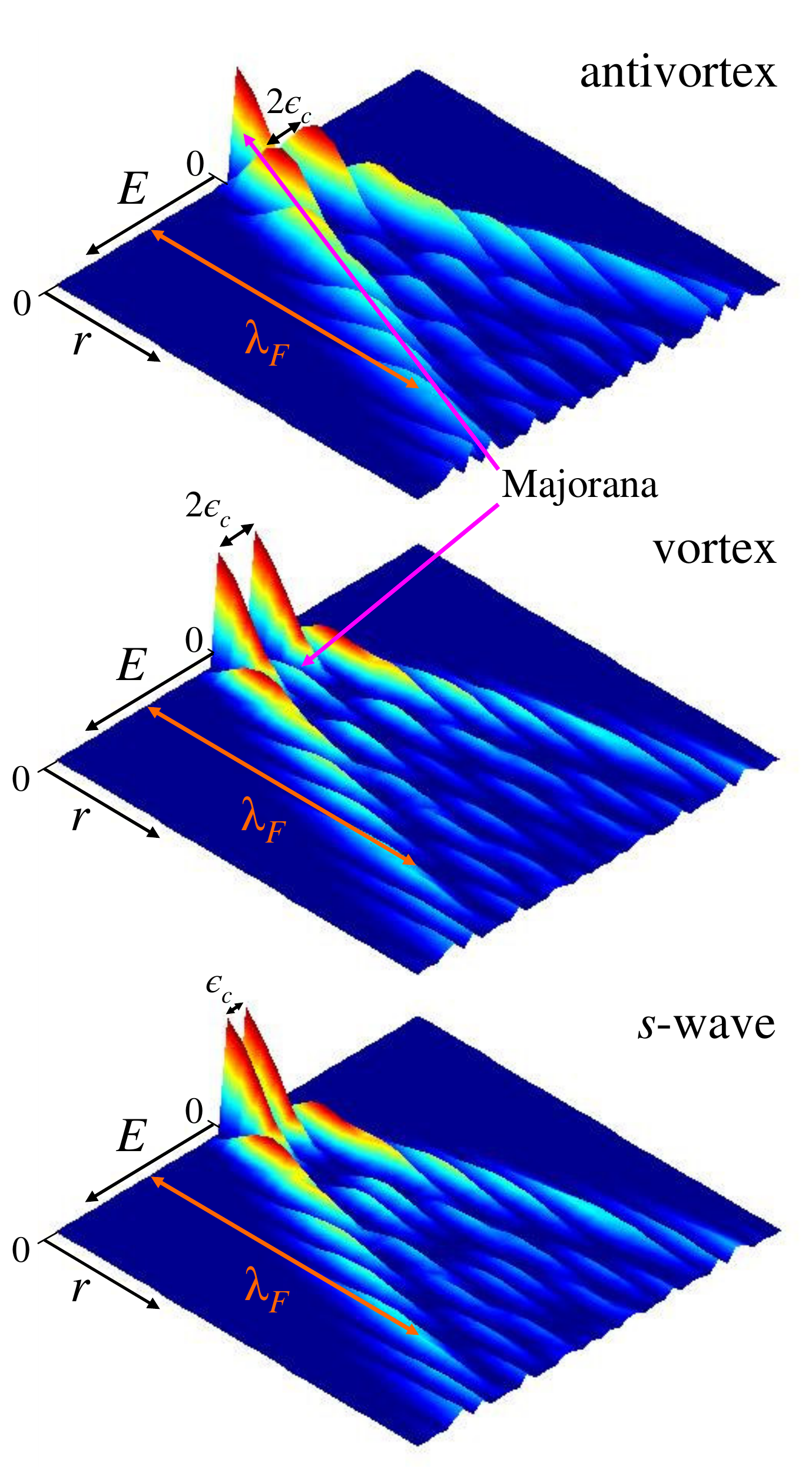}
\caption{   \label{Fig:LDOS} %
Zero temperature LDOS ${\cal T}(E,r)$, in the cores of the \pxipy
antivortex (top) and vortex (middle), and of the \Swave vortex
(bottom). $\eC$ is the spacing between CdGM states, and $\lambda_F$
is the Fermi wavelength. The peaks belong to the CdGM states. Notice
that the zero energy Majorana state is maximized at the origin in
the \pxipy antivortex, while it is removed from the origin in the
vortex, and is absent in the \Swave. }
\end{center}
\end{figure}

At zero temperature the tunneling local density of states (LDOS) is
defined as \cite{Shore}
\begin{equation}   \label{Eq:LDOS}
    {\cal T}(E,r) = \sum_{n} |u_n(r)|^2 \delta(E-E_n) + |v_n(r)|^2 \delta(E+E_n),
\end{equation}
where $r$ is the distance from the vortex (or antivortex) center.
Fig.~\ref{Fig:LDOS} shows the LDOS at the core, for displacements $r
\le 0.3\xi$ and energies $|E|\le 0.2\Delta_0$. The \pxipy state
shows a distinction between the antivortex and the vortex, while in
the \Swave they are the same.

The Majorana state can be easily discerned as zero energy peaks near
the centers of the vortex and antivortex cores of the \pxipy
superconductor. The other CdGM core states also appear as
oscillatory peaks, with energy spacing $\eC$. The difference between
the vortex and antivortex of the \pxipy LDOS is apparent: according
to Eq.~(\ref{Eq:u0}) the antivortex Majorana state wavefunction
behaves like $J_0$, which is peaked at $r = 0$. By contrast, the
vortex Majorana state wavefunction behaves like $J_1$, which is zero
at $r = 0$, and has a smaller peak at $r \approx \lambda_F/4$, where
$\lambda_F = 2\pi/k_F$ is the Fermi wavelength.

The origin of this distinction was discussed in Sec.~\ref{Sec:BdG}.
According to that argument, in the antivortex only the wavefunction
of the Majorana state, with $m = \half$, is real and peaked at the
origin. All the excitations with $m \neq \half$ have a phase
winding, and are therefore equal to zero at the center of the
vortex, and have much lower peaks. On the other hand, in the vortex
the Majorana state wavefunction has a phase winding of $e^{i\phi}$,
and must vanish at the origin. But the $u_m(\xvec)$ part of the
first excited state, with $m = -\half$ and $E_{1/2}^c = \eC$, is
real and finite at the center of the vortex. Similarly the
$v_m(\xvec)$ part of its negative companion, with $m = {3 \over 2}$
and $E_{3/2}^c = -\eC$, is also real and finite at the center of the
vortex. Therefore in the LDOS the first two excitations have
relatively large peaks at $r = 0$, while all the other CdGM states
are much lower, as seen in Fig.~\ref{Fig:LDOS}.

The same argument holds for the \Swave superconductor. The phase
winding of the order parameter with a vortex is $\e^{i\phi}$, while
with an antivortex it is $\e^{-i\phi}$. This sign of the phase
winding is only a matter of convention; there is no distinction in
the \Swave superconductor between the vortex and the antivortex.
Physically speaking, the pairing of the particles does not involve
any internal angular momentum, so the sign of the vorticity is
meaningless. For such an order parameter there is no Majorana
fermion solution, since it requires $u_0(\xvec) = \tilde{u}_0(r)
\e^{i\phi/2}$, which is not single valued. However, for the lowest
state, with $m = 0$ and $E_0^c = \half\eC$, $u_m(\xvec)$ is real.
Similarly $v_m(\xvec)$ is real for $m = 1$ (and $E_1^c =
-\half\eC$). Therefore, the two lowest states are peaked at the
origin, while all the higher excitations are equal to zero at the
origin, and therefore much lower, in agreement with
Fig.~\ref{Fig:LDOS}.

In a tunneling spectroscopy experiment (e.g.~Ref.~\cite{Davis}), the
tunneling conductance is measured. The conductance reflects the
smearing of the LDOS by temperature broadening \cite{Gygi}
\begin{equation}  \label{Eq:dIdV}
    {\mathrm{d}I \over \mathrm{d}V}(E,r) \sim
    T \int \mathrm{d}E' \left({ \partial f(E-E') \over \partial E' }\right) {\cal T}(E',r),
\end{equation}
where $f(E)$ is the Fermi-Dirac distribution at zero chemical
potential and temperature $T$. In the BCS weak coupling regime,
$\Delta_0 \ll \eF$, and therefore $\eC$ could be a very small
temperature scale. At moderate temperatures $\eC < T < \Delta_0$,
the peaks of Fig.~\ref{Fig:LDOS} are smeared on the energy axis (but
not on the $r$ axis).

\begin{figure}[htb]
\begin{center}
\includegraphics[width=7.5cm,angle=0]{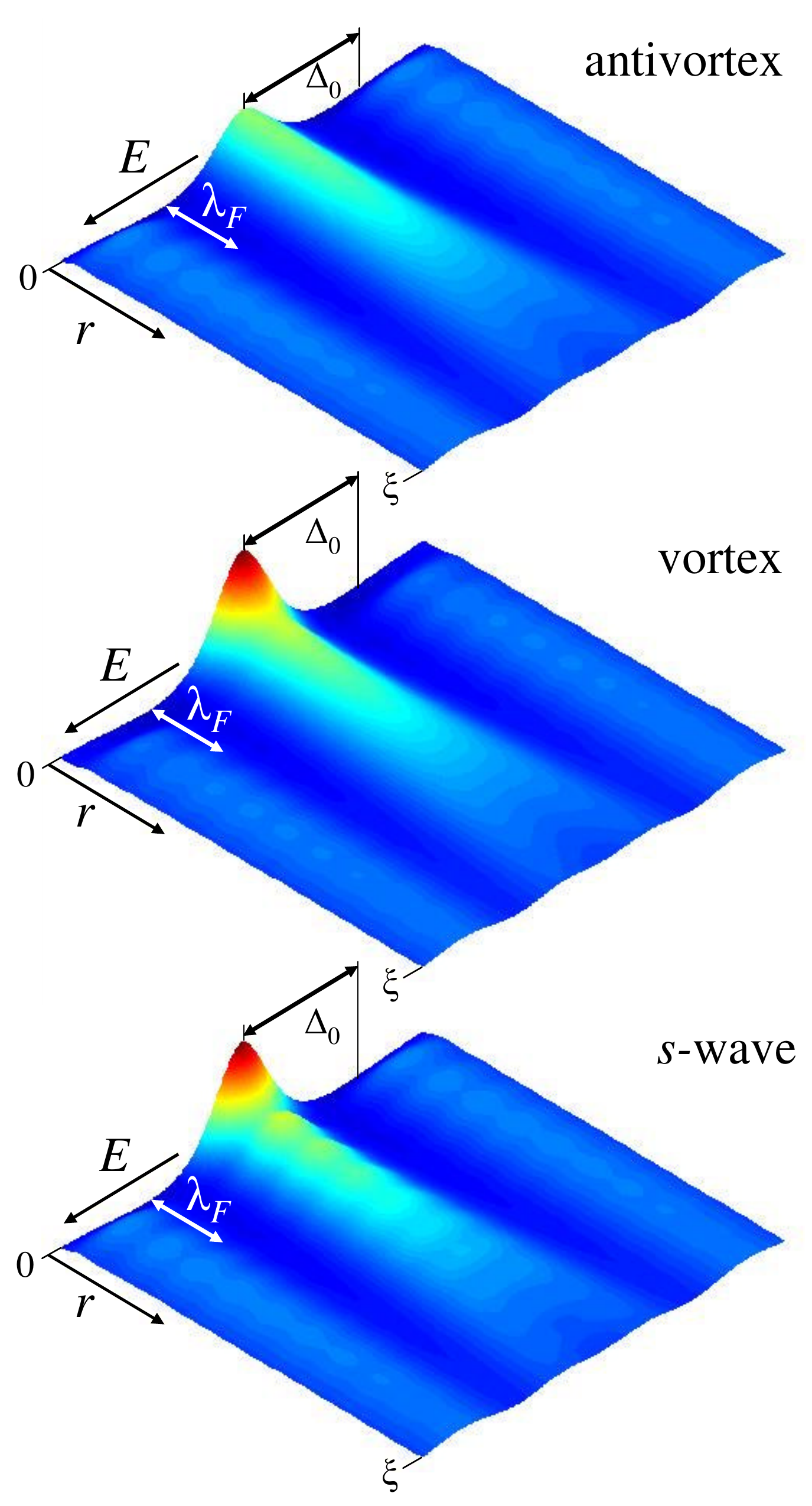}
\caption{    \label{Fig:dIdV} %
Tunneling conductance ${\mathrm{d}I \over \mathrm{d}V}(E,r)$, in
arbitrary units, of the \pxipy antivortex (top) and vortex (middle),
and of the \Swave vortex (bottom). $\Delta_0$ is the energy gap, and
$\xi$ is the coherence length, which specifies the radiuses of the
cores. The temperature \mbox{$T = 0.15 \Delta_0$}, is about 10 times
larger than the CdGM level spacing. The conductance of the \pxipy
vortex is almost identical to that of the \Swave, while the central
peak of the antivortex is twice lower. }
\end{center}
\end{figure}

The tunneling conductance at $T = 0.15 \Delta_0 = 7.5\, \eC$, for
displacements $r \le \xi$ and energies $|E| \le 1.5\Delta_0$, is
depicted in Fig.~\ref{Fig:dIdV} for the \pxipy antivortex and
vortex, and for the \Swave vortex. The conductance shows a central
peak at $E = 0, r = 0$, with low broad ridges dispersing away to
larger $E, r$. We see that while the conductance of the \pxipy
vortex is almost identical to the conductance of the \Swave vortex,
\emph{the central peak of the antivortex is half the height of that
of the vortex}.

\begin{figure}[htb] \vspace{0cm}
\begin{center}
\includegraphics[width=8cm,height=5cm,angle=0]{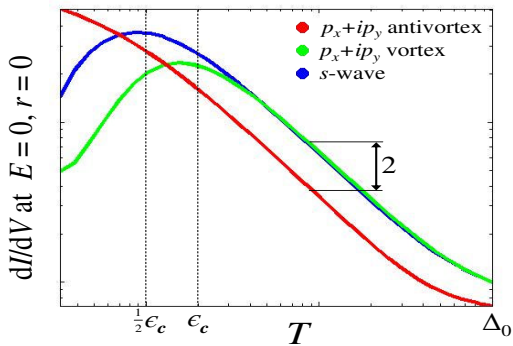}
\caption{   \label{Fig:peak} %
Zero bias conductance peak at the (anti)vortex core as a function of
temperature $T$, in log-log scale. For $\eC < T < \Delta_0$ the
\pxipy vortex (green) and \Swave (blue) peaks are twice the height
of the antivortex peak (red).}
\end{center}
\end{figure}

The explanation for this factor of 2 comes from the origin of the
zero bias peak of the conductance. Consider the conductance of the
\Swave vortex. Since this lacks a zero energy CdGM state, the zero
bias conductance vanishes at $ T \ll \eC$. When the temperature is
raised to $\half\eC$, the two lowest CdGM states, which are peaked
at the origin for $E \approx \pm\half\eC$, broaden to create a zero
bias peak. As the temperature rises further, this peak diminishes
due to the broadening, as shown by the blue curve of
Fig.~\ref{Fig:peak}.

The \pxipy vortex behaves similarly to the \Swave case, as shown by
the green curve of Fig.~\ref{Fig:peak}. For $ T \ll \eC$ the zero
bias conductance peak vanishes, since the Majorana state
wavefunction has a node at the origin. By contrast, for $ T \geq
\eC$ a zero bias peak is created by the two broadened first excited
states, which are peaked at the origin with $E \approx \pm\eC$. On
the other hand, in the antivortex the Majorana state wavefunction is
peaked exactly at the origin. Therefore its conductance is maximal
at $T = 0$, and diminishes with rising temperature, as depicted by
the red curve of Fig.~\ref{Fig:peak}. Furthermore, since in the
antivortex {\em only} the Majorana state is peaked at the origin,
the height of the zero bias peak of the antivortex is determined by
{\em the broadening of a single state}, while in the vortex and in
the \Swave it is {\em a sum of the broadening of two states.} This
is why the central peak in the antivortex is half the height of that
found in the vortex, when $\eC < T < \Delta_0$. Thus, the asymmetry
between the vortex and the antivortex is not only a clear
fingerprint of the \pxipy symmetry of the order parameter, but it is
also a smoking gun evidence of the existence of the Majorana state
itself.

\begin{figure}[htb]
\begin{center}
\includegraphics[width=7.5cm,height=4.5cm,angle=0]{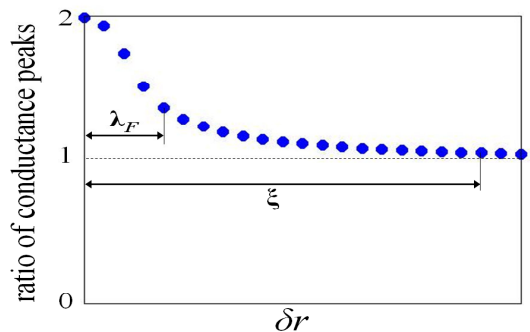}
\caption{   \label{Fig:ratio} %
Effect of spatial resolution. Ratio of vortex to antivortex
conductance peak heights (shown in Fig.~\ref{Fig:peak} for perfect
resolution), for spatial resolution $\delta r$. $\lambda_F$ and
$\xi$ are Fermi wavelength and coherence lengths respectively.
Temperature is $7.5\, \eC$.  }
\end{center}
\end{figure}

The distinction between the vortex and the antivortex spectra is
based on the distinction between the wavefunctions, which may be
seen on scale of the Fermi wavelength $\lambda_F$. Therefore, as
long as the spatial resolution in the tunneling conductance $\delta
r$ is better than $\lambda_F$, our effect is observable. But when
${\mathrm{d}I \over \mathrm{d}V}(E,r)$ (Eq.~\ref{Eq:dIdV}) is
smeared over a length scale $(\delta r)^2 > \lambda_F^2$, the ratio
between the vortex and antivortex peak heights rapidly approaches
unity as $\delta r > \lambda_F$, as shown in Fig.~\ref{Fig:ratio}.
The requirement of such a high resolution, in spite of being
restrictive, is feasible in present technology of $\delta r \approx
1\textrm{\AA}$, since in \SRO \ $\lambda_F \approx 8.3\textrm{\AA}$
\cite{Maeno}.

Moreover, although we found in Sec.~\ref{Sec:Disorder} that the
Majorana state survives moderate disorder, the suggested results
hold provided the disorder is weak on the scale of $\lambda_F$,
which is very reasonable as that limit corresponds to minimum
conductivity and Anderson localization.

The above analysis is based on the two-dimensionality of the sample.
In real three-dimensional samples the conductance peaks are strongly
suppressed by bulk states and surface imperfections
\cite{Davis,Hess}. Nevertheless, it is the {\em difference} between
a vortex peak and an antivortex peak which is sensitive to the
existence of the Majorana state. Moreover, according to the
asymptotic analysis \cite{KopninSalomaa}, momentum along the vortex
line (additional $e^{ik_z z}$ in the wavefunction), leaves the
dispersion of the core states (Eq.~\ref{Eq:Emc}) unchanged, and only
modifies weakly the oscillations in the radial part of the
wavefunction (through the oscillatory argument, which becomes
$\sqrt{k_F^{\phantom{w}2} + k_z^{\phantom{c}2}}\,r$).

Performing such an experiment is certain to be a challenge, although
likely remains possible. One might imagine leaving the tip of the
STM in the same position while reversing a weak magnetic field where
the field is weak enough so as to flip the direction of the vortex,
but not to overturn the chiral order parameter. Such an experiment
would rely on an assumption that the vortex prefers to sit at one
particular position in the sample
--- presumably due to some inhomogeneity or disorder in the sample
--- and that this preferred position does not change when the weak
magnetic field is reversed. In practice, however, one would have to
check this assumption by spatially scanning the STM as even a small
change in the vortex position would have a large effect on the
measured tunneling. \\


{\em General Cases.} A difference between vortex and antivortex
conductance peak is expected for any chiral symmetry breaking (CSB)
superconductor. The important questions are whether this difference
is observable for $T > \eC$, and whether it is sensitive to the
existence of Majorana fermions.

For a CSB superconductor with relative angular momentum $M = 1, 2,
\ldots$ (chiral-{\it p}, chiral-{\it d}, $\ldots$), in the presence
of a vortex with vorticity $N = \pm1, \pm2, \ldots$, the phase
winding of the order parameter is $\e^{i(M + N)\phi}$. A Majorana
fermion solution $u_0 \sim \e^{im\phi}$ requires $m = (M + N)/2$,
and would be single valued only if $m$ is an integer, i.e. $M + N$
is even. However, only if $M + N = 0$ can the wavefunction of the
Majorana state be finite at the center of the vortex, i.e. an
antivortex with vorticity $N = -M$. Therefore, although a Majorana
state is expected for every even $M + N$ \cite{Volovik}, the
asymmetry in the zero bias conductance peak for $ T > \eC$ is
expected only in the subset $|N| = M$. Notice, that the \pxipy
vortex is the only case where the asymmetry occurs for a unit
vorticity.


\section{Summary}
\label{Sec:Summary} %
Implementing the BdG equation of a \pxipy superconductor on a sphere
had several advantages. It allowed us to study the system with and
without vortices and edges, without artifacts of finite size
boundaries. We found the exponential decay of the Majorana state
energy with respect to the intervortex distance to be robust with
respect to the addition of moderate disorder strength, up to the
Fermi energy scale. The tunneling conductance was plotted, and
compared to the \Swave case. It was found that in the moderate
temperature regime $\eC < T < \Delta_0$, the zero bias conductance
in the center of the \pxipy antivortex is half the height of that of
the vortex. The asymmetry in the conductance peak heights is a
smoking gun signature of the spin polarized \pxipy superconductor,
and a direct measurement of the existence of the Majorana state.
This idea can be generalized to order parameters of higher angular
momentum.


\section*{ACKNOWLEDGEMENTS}
We thank Ady Stern for useful discussions. Support from US - Israel
Binational Science foundation and Israel Science foundation is
acknowledged. AA acknowledges Aspen Center for Physics for its
hospitality. HAF acknowledges the support of the NSF through Grant
No. DMR-0704033. AA and SHS acknowledge the hospitality of the KITP,
where this collaboration was initiated.

\appendix
\section{The matrix element $\Delta^V_{lm,l'm'}$}
\label{App:DVlmlm} %
In this appendix we derive $\Delta^V_{lm,l'm'}$
(Eqs.~\ref{Eq:DVlmlm}-\ref{Eq:Dl}) from $\Delta_V$
(Eqs.~\ref{Eq:Delta_V}-\ref{Eq:ab}).

The spinor functions $\alpha$ and $\beta$ (Eq.~\ref{Eq:ab}) obey
\begin{eqnarray}  \label{Eq:ab_ab}
    \alpha \beta' - \beta \alpha' & = &
        2\pi\left(
        Y_{-\half \half  \half}(\omgvec) Y_{-\half \half -\half}(\omgvec') \right. \nonumber \\
    & & \phantom{2\pi} \left.
        - Y_{-\half \half -\half}(\omgvec) Y_{-\half \half  \half}(\omgvec') \right),
\end{eqnarray}
and
\begin{eqnarray}  \label{Eq:aa_bb_1}
    && | \alpha \alpha'^* + \beta \beta'^* |^{2\Rxi} \nonumber \\
    && \qquad = \left| \cos{\theta \over 2} \cos{\theta' \over 2} +
                       \sin{\theta \over 2} \sin{\theta' \over 2} \e^{i(\phi - \phi')}
                \right|^{2\Rxi} \nonumber \\
    && \qquad = \left( \frac{1 + \cos\theta \cos\theta' + \sin\theta \sin\theta'
                                   \cos(\phi-\phi')}{2} \right)^{\Rxi} \nonumber \\
    && \qquad = \left( \frac{1 + \cos\Theta}{2} \right)^{\Rxi},
\end{eqnarray}
where $\Theta$ is the angle between $\omgvec$ and $\omgvec'$. Note
that for $\xi_p \ll R$
\begin{equation} \label{Eq:cos_e}
    \left( \frac{1 + \cos\Theta}{2} \right)^{\Rxi}
    \approx \e^{-{\Rxi \over 4} \Theta^2},
\end{equation}
which is equivalent to Gaussian pairing in the plane.

If we denote $M = \left[ \Rxi \right]$, then
\begin{eqnarray}  \label{Eq:aa_bb_2}
    | \alpha \alpha'^* + \beta \beta'^* |^{ 2\Rxi }
    & \approx & {1 \over 2^M} \left( 1 + \cos\Theta \right)^M \\
    & = & \frac{1}{2^M} \sum_{n=0}^M \binom{M}{n} \cos^n\Theta \nonumber.
\end{eqnarray}
Any monomial $x^n$ can be expressed as a series of Legendre
polynomials $P_l(x)$ \cite{Abramowitz}
\begin{equation}  \label{Eq:xn_Pl}
    x^n = \sum_{l=n,n-2,...} \frac{(2l+1)n!}{2^\frac{n-l}{2} ({n-l \over 2})!
          (l+n+1)!!} P_l(x),
\end{equation}
which satisfy \cite{Edmonds}
\begin{equation}   \label{Eq:Pl_Ylm}
    P_l(\cos\Theta) = \frac{4\pi}{2l + 1}
                      \sum_{m=-l}^{l} (-1)^m Y_{l-m}(\omgvec) Y_{lm}(\omgvec') ,
\end{equation}
where $Y_{lm}$ ($Y_{0lm}$) are the spherical harmonics. Substituting
Eqs.~(\ref{Eq:xn_Pl}) and (\ref{Eq:Pl_Ylm}) into
Eq.~(\ref{Eq:aa_bb_2}), we obtain
\begin{eqnarray}  \label{Eq:aa_bb_3}
    && | \alpha \alpha' + \beta \beta' |^{ 2\Rxi } \approx \\
    && \qquad 4\pi \frac{M!}{2^M} \sum_{l=0}^{M} \sum_{m=-l}^{l} (-1)^m A_l^M
                                Y_{l-m}(\omgvec) Y_{lm}(\omgvec'),  \nonumber \\
    && A_l^M = \sum_{n=l,l+2,\ldots}^{M}
               \left[ 2^\frac{n-l}{2} \left( \frac{n-l}{2} \right)!
               \left( M-n \right)! \left( l+n+1 \right)!! \right]^{-1}.  \nonumber
\end{eqnarray}

Multiplying two harmonics yields \cite{WuYang}
\begin{eqnarray} \label{Eq:YqlmYqlm}
    && Y_{qlm}(\omgvec)Y_{q'l'm'}(\omgvec) = (-1)^{l+l'-q-q'-m-m'} \\
    && \qquad \times \sqrt{ {\textstyle {1 \over 4\pi} } (2l+1) (2l'+1) }
       \sum_{l'' = |l-l'|}^{l+l'} (-1)^{l''} \sqrt{2l''+1} \nonumber \\
    && \qquad \qquad \times \Wj{l}{l'}{l''}{m}{m'}{-m-m'}
                            \Wj{l}{l'}{l''}{q}{q'}{-q-q'} \nonumber \\
    && \qquad \qquad \times Y_{q+q',l'',m+m'}(\omgvec). \nonumber
\end{eqnarray}
In particular
\begin{eqnarray} \label{Eq:Y0lmY1_2}
    && Y_{0lm}(\omgvec)Y_{-\half \half \pm\half}(\omgvec) = (-1)^{m-1} \nonumber \\
    && \qquad \times \sqrt{ {\textstyle {1 \over 2\pi} } (2l+1) }
       \sum_{s = \pm\half} (-1)^{s \mp \half} \sqrt{ 2l'+1 \pm 2s } \nonumber \\
    && \qquad \qquad \times \Wj{l}{\half}{l+s}{m}{\pm\half}{-m\mp\half}
                            \Wj{l}{\half}{l+s}{0}{-\half}{\half} \nonumber \\
    && \qquad \qquad \times Y_{-\half,l+s,m\pm\half}(\omgvec) \nonumber \\
    && \qquad = \sqrt{ {\textstyle {1 \over 4\pi} } } \left( \sqrt{ \frac{l \pm m+1}{2l+1} }
                 Y_{-\half,l+\half,m\pm\half}(\omgvec) \pm \right. \nonumber \\
    && \qquad \qquad \qquad \left. \sqrt{ \frac{l \mp m+1}{2l+1} }
                 Y_{-\half,l-\half,m\pm\half}(\omgvec) \right),
\end{eqnarray}
where the last expression comes from writing the 3j symbols
explicitly \cite{Edmonds}.

Substituting Eqs.~(\ref{Eq:ab_ab}), (\ref{Eq:aa_bb_3}) and
(\ref{Eq:Y0lmY1_2}) into Eq.~(\ref{Eq:Delta_p}) yields
\begin{eqnarray}  \label{Eq:Dp_lmlm}
    && \Delta_p (\omgvec,\omgvec') =
       \frac{\Delta_0}{R^2} \frac{M^2 \cdot M!}{ 2^{M + 1} (l_F + \half) }
       \sum_{l = 0}^{M} ( A_l^M - A_{l+1}^M ) \nonumber \\
    && \qquad \times \sum_{m = -l-1}^{l} (-1)^{m + 1}
       Y_{-\half,l+\half,-m-\half}(\omgvec) Y_{-\half,l+\half,m+\half}(\omgvec')  \nonumber \\
    && \qquad = \frac{\Delta_0}{R^2} \sum_{l = \half}^{M + \half} D_l \\
    && \qquad \qquad \times \sum_{m = -l}^{l} (-1)^{m - \half}
       Y_{-\half lm}(\omgvec) Y_{-\half l-m}(\omgvec'),  \nonumber
\end{eqnarray}
with
\begin{eqnarray}  \label{Eq:Dl_M}
    && D_l = \frac{M^2 \cdot M!}{ 2^{M + 1} (l_F + \half) } \bigg( B_l^M +  \\
    && \qquad \qquad \sum_{n = 0,2,\ldots}^{M - l - \half}
       \left( \frac{1}{M - l + \half - n} - \frac{1}{2l + 2 + n} \right) \nonumber \\
    && \qquad \qquad \times \left[ 2^{n \over 2} \left( {n \over 2} \right)!
       (M - l - \half - n)! (2l + n)!! \right]^{-1} \bigg), \nonumber
\end{eqnarray}
where $B_l^M = \left[ 2^{ \half \left( M - l + \half \right) }
        \left( \half \left( M - l + \half \right) \right)!
        \left( M + l + \half \right)!! \right]^{-1}$
only for $l - \half \equiv M \; (mod\;2)$.

Alternatively, substituting $k_x + ik_y \rightarrow (l + \half)/R$
in the planar $\Delta_{\bf k}$ (Eq.~\ref{Eq:Dk}), yields an
excellent approximation for $D_l$
\begin{equation}  \label{Eq:Dl_app}
    D_l \approx \frac{l + \half}{l_F + \half}
                \e^{-(l + \half)^2 \xiR },
\end{equation}
which is valid for $R > 5\xi_p$. For $l$, $l_F \gg 1$ we obtain
Eq.~(\ref{Eq:Dl}).

For short range pairing $\xi_p \ll R$, the vorticity of the center
of mass can be approximated by
\begin{equation}  \label{Eq:FV_app}
    F_V(\bar{\omgvec}) \approx
        {\textstyle \half} \big( F_V(\omgvec) + F_V(\omgvec') \big).
\end{equation}
$F_V$ is expanded in spherical harmonics,
\begin{equation}  \label{Eq:fVL}
    F_V(\omgvec) = \sum_{L=1,3,5,\ldots} f^V_L Y_{L1}(\omgvec),
\end{equation}
which defines $f^V_L$.

Substituting Eqs.~(\ref{Eq:Dp_lmlm}) and (\ref{Eq:fVL}) into
Eq.~(\ref{Eq:Delta_V}), and using Eq.~(\ref{Eq:YqlmYqlm}), yield
\begin{eqnarray}
    && \Delta_V (\omgvec,\omgvec') = \frac{\Delta_0}{R^2}
       \sum_{ \begin{subarray}{c} lm \\ l'L \end{subarray} }
       (-1)^{l + l' + L}  D_l \cdot f^V_L \nonumber \\
    && \qquad \times \sqrt{ {\textstyle {1 \over 16\pi} } (2l+1) (2l'+1) (2L+1) }
       \Wj{l}{L}{l'}{-\half}{0}{\half} \nonumber \\
    && \qquad \times \left( \Wj{l}{L}{l'}{-m}{1}{m-1}
       Y_{-\half lm}(\omgvec) Y_{-\half l'-m+1}(\omgvec') \right. \nonumber \\
    && \qquad \qquad \left. - \Wj{l}{L}{l'}{m}{1}{-m-1}
       Y_{-\half l'm+1}(\omgvec) Y_{-\half l-m}(\omgvec') \right) \nonumber \\
    && \qquad = {1 \over R^2} \sum_{lml'm'} \Delta^V_{lm,l'm'} Y_{-\half lm}(\omgvec)
       Y_{-\half l'm'}(\omgvec').
\end{eqnarray}
The last line defines $\Delta^V_{lm,l'm'}$. Finally, by using
\cite{Edmonds}
\begin{eqnarray}  \label{Eq:W3j}
    \Wj{l}{l'}{l''}{m}{m'}{m''}
        & = & \Wj{l''}{l}{l'}{m''}{m}{m'} \\
        & = & (-1)^{l+l'+l''} \Wj{l''}{l'}{l}{m''}{m'}{m} \nonumber \\
        & = & (-1)^{l+l'+l''} \Wj{l}{l'}{l''}{-m}{-m'}{-m''}, \nonumber
\end{eqnarray}
we obtain
\begin{eqnarray}  \label{Eq:Dl_L}
    \Delta^V_{lm,l'm'} & = & \delta_{m', -m+1} \Delta_0
       \sqrt{ {\textstyle {1 \over 16\pi} } (2l + 1) (2l' + 1) } \\
    && \times \sum_{L} \sqrt{ 2L + 1 } \; f^V_L
       \left( D_l - (-1)^{l + l' + L} D_{l'} \right) \nonumber\\
    && \qquad \times \Wj{l}{l'}{L}{\half}{-\half}{0}
                     \Wj{l}{l'}{L}{-m}{m-1}{1}. \nonumber
\end{eqnarray}
Since $L$ is odd, we obtain Eq.~(\ref{Eq:DVlmlm}). Note that
$\Delta^V_{l,m,l',-m+1} = -\Delta^V_{l',-m+1,l,m}$, as expected by
the antisymmetry.

\section{The disorder potential}
\label{App:Wnoise} %
The white noise potential is defined as a series of independently
identically distributed complex elements $w_{lm}$, with both real
and imaginary parts uniformly distributed in the interval $\left[
-\frac{\sqrt{ 6\pi }\w}{\lL}, \frac{\sqrt{ 6\pi }\w}{\lL} \right]$
for every $l \leq \lL$. Therefore
\begin{eqnarray}  \label{Eq:wlmwlm}
    \langle w_{lm}^* w_{l'm'} \rangle
        & = & \delta_{ll'} \delta_{mm'} \langle |w_{lm}|^2 \rangle \\
        & = & \delta_{ll'} \delta_{mm'} 2 \langle |Re(w_{lm})|^2 \rangle
            \nonumber \\
        & = & \delta_{ll'} \delta_{mm'} \frac{4\pi}{\lL^{\phantom{0}2}} \ww
            \qquad \quad \forall l \leq \lL. \nonumber
\end{eqnarray}
In real space
\begin{eqnarray} \label{Eq:WW}
    \langle W^2(\omgvec) \rangle
        & = & \sum_{ \begin{subarray}{c} l,m \\ l',m' \end{subarray} } ^{\lL}
              \langle w_{lm}^* w_{l'm'} \rangle Y_{lm}^*(\omgvec) Y_{l'm'}(\omgvec)
              \nonumber \\
        & = & \ww \frac{4\pi}{\lL^{\phantom{0}2}} \sum_{l,m}^{\lL} |Y_{lm}(\omgvec)|^2.
\end{eqnarray}
In particular for the north pole
\begin{eqnarray} \label{Eq:WW_N}
    \langle W^2(\theta = 0) \rangle
        & = & \ww \frac{4\pi}{\lL^{\phantom{0}2}} \sum_{l=0}^{\lL} |Y_{l0}(\theta=0)|^2
              \nonumber \\
        & = & \ww \frac{4\pi}{\lL^{\phantom{0}2}} \sum_{l=0}^{\lL} \frac{2\lL + 1}{4\pi}
              \nonumber \\
        & \approx & \ww.
\end{eqnarray}\\

In order to calculate the matrix element $W_{lm,l'm'}$, we use the
identities \cite{WuYang}
\begin{equation}  \label{Eq:Yqlm_1}
    Y_{qlm}^*(\omgvec) = (-1)^{q+m} Y_{-ql-m}(\omgvec),
\end{equation}
and
\begin{eqnarray}  \label{Eq:YqlmYqlmYqlm}
    && \int \mathrm{d}\omgvec \, Y_{qlm}(\omgvec) Y_{q'l'm'}(\omgvec)
       Y_{q''l''m''}(\omgvec) = \nonumber \\
    && \qquad (-1)^{l+l'+l}
       \sqrt{ {\textstyle {1 \over 4\pi} } (2l+1) (2l'+1) (2l''+1) }
       \nonumber \\
    && \qquad \times \Wj{l}{l'}{l''}{q}{q'}{q''} \Wj{l}{l'}{l''}{m}{m'}{m''}.
\end{eqnarray}
Hence
\begin{eqnarray}  \label{Eq:Wlmlm2}
    W_{lm,l'm'} & = & \int \mathrm{d}\omgvec Y_{-\half lm}^*(\omgvec)
                      W(\omgvec) Y_{-\half l'm'}(\omgvec) \\
                & = & (-1)^{m - \half} \sum_{l'',m''} w_{l''m''} \nonumber \\
                &   & \quad \times \int \mathrm{d}\omgvec Y_{\half l-m}(\omgvec)
                      Y_{-\half l'm'}(\omgvec) Y_{0l''m''}(\omgvec), \nonumber
\end{eqnarray}
which yields Eq.~(\ref{Eq:Wlmlm}). Note that $W_{l'm',lm} =
W_{lm,l'm'}^*$, as expected from a real potential.

\section{Matrix elements of edge operators}
\label{App:MatrixE} %
Both the edge potential operator $W_E$ (Eq.~\ref{Eq:WE}) and order
parameter $\Delta_E$ (Eq.~\ref{Eq:DE}) have the following matrix
elements:
\begin{eqnarray}
    W^E_{lm,l'm'} & = & \delta_{m',-m} (-1)^{m - \half} \sqrt{ {\textstyle {1 \over 4\pi} }
                        (2l + 1) (2l' + 1) }  \nonumber \\
                  &   & \times \sum_{L} \sqrt{2L + 1} \; w^E_L  \label{Eq:WElmlm}\\
                  &   & \qquad \times \Wj{l}{l'}{L}{\half}{-\half}{0}
                                      \Wj{l}{l'}{L}{m}{-m}{0},   \nonumber \\
    \Delta^E_{lm,l'm'} & = & \delta_{m',-m} \Delta_0 \sqrt{ {\textstyle {1 \over 16\pi} }
                        (2l + 1) (2l' + 1) }  \label{Eq:DElmlm} \\
                  &   & \times \sum_{L} \sqrt{2L + 1} \; f^E_L
                        \left( D_{l'} - (-1)^{l + l' + L} D_l \right)  \nonumber \\
                  &   & \qquad \times \Wj{l}{l'}{L}{\half}{-\half}{0}
                                      \Wj{l}{l'}{L}{m}{-m}{0}  \nonumber.
\end{eqnarray}
$W^E_{lm,l'm'}$ is obtained by substituting $m = m'$ in
Eq.~(\ref{Eq:Wlmlm}) (due to Eq.~\ref{Eq:WE}). $\Delta^E_{lm,l'm'}$
is essentially Eq.~(\ref{Eq:Dl_L}). But since $F_E$
(Eq.~\ref{Eq:fE}) replaces $F_V$ (Eq.~\ref{Eq:fVL}), $m$ is coupled
to $-m$, and $f^E_L \neq 0$ for both odd and even $L$'s.

For the case of an edge with a vortex, the matrix elements of the
order parameter $\Delta_{VE}$ (Eq.~\ref{Eq:DVE}) are
\begin{eqnarray}
    \Delta^{V\!E}_{lm,l'm'} & = & \delta_{m',-m+1} \Delta_0 \sqrt{ {\textstyle {1 \over 16\pi} }
                        (2l + 1) (2l' + 1) }  \label{Eq:fVElmlm} \\
                  &   & \times \sum_{L} \sqrt{2L + 1} \; f^{V\!E}_L
                        \left( D_{l} - (-1)^{l + l' + L} D_l' \right)  \nonumber \\
                  &   & \qquad \times \Wj{l}{l'}{L}{\half}{-\half}{0}
                                      \Wj{l}{l'}{L}{-m}{m-1}{1}  \nonumber,
\end{eqnarray}
which are exactly Eq.~(\ref{Eq:Dl_L}), with $f^{V\!E}_L \neq 0$ for
both odd and even $L$'s.


\end{document}